\documentclass[11pt,english]{article}
\pdfoutput=1
\usepackage[a4paper,bindingoffset=0.2in,%
left=1in,right=1in,top=1in,bottom=1in,%
footskip=.25in]{geometry}
\usepackage[utf8]{inputenc}
\usepackage{jcappub}
\date{}
\usepackage{fixltx2e}
\usepackage{xcolor}
\usepackage{mathrsfs}
\usepackage{amsmath}
\usepackage{amssymb}
\usepackage{bm}
\usepackage{url}
\usepackage{wrapfig}
\usepackage[colorinlistoftodos]{todonotes}
\usepackage{graphicx}
\usepackage{epsfig}
\usepackage{physics}
\usepackage{soul}
\usepackage[titletoc]{appendix}
\usepackage[autostyle=true]{csquotes} 
\usepackage[normalem]{ulem}
\usepackage{framed}
\usepackage{tensor}
\usepackage{caption}
\usepackage{subcaption}
\usepackage{float}
\usepackage{framed}
\usepackage{xparse}
\usepackage{cancel}

\NewDocumentCommand{\evalat}{sO{\big}mm}{%
	\IfBooleanTF{#1}
	{\mleft. #3 \mright|_{#4}}
	{#3#2|_{#4}}%
}
\newcommand{\kmpc}{\ensuremath{\text{ Mpc}^{-1}}}

\newcommand{\qop}[1]{\left\langle{#1}\right\rangle}

\newcommand{\rhogw}{{\rho_{{\text{\tiny{GW}}}}}}
\newcommand{\Omegagw}{{\Omega_{{\text{\tiny{GW}}}}}}

\newcommand{\deltagw}{\ensuremath{\delta^{\text{\tiny{GW}}}}}
\newcommand{\omegagw}{\ensuremath{\omega_{\text{\tiny{GW}}}}}
\newcommand{\deltaT}{\ensuremath{\delta^{\text{\tiny T}}}}
\newcommand{\GT}{\ensuremath{C^{\text{\tiny GW-T}}_{\ell}}}
\newcommand{\GG}{\ensuremath{C^{\text{\tiny GW}}_{\ell}}}
\newcommand{\TT}{\ensuremath{C^{\text{\tiny TT}}_{\ell}}}
\newcommand{\vk}{\ensuremath{\vec{k}}}
\newcommand{\vq}{\ensuremath{\vec{q}}}
\newcommand{\vp}{\ensuremath{\vec{p}}}
\newcommand{\vx}{\ensuremath{\vec{x}}}
\newcommand{\hk}{\ensuremath{\hat{k}}}
\newcommand{\hq}{\ensuremath{\hat{q}}}
\newcommand{\hp}{\ensuremath{\hat{p}}}
\newcommand{\hx}{\ensuremath{\hat{x}}}
\newcommand{\hn}{\ensuremath{\hat{n}}}
\newcommand{\hv}{\ensuremath{\hat{v}}}

\newcommand{\fnl}{\ensuremath{F_{\rm NL}}}

\DeclareMathOperator{\sinc}{sinc}
\newcommand{\tj}[6]{\ensuremath{\begin{pmatrix}
			#1 & #2 & #3 \\
			#4 &#5 &#6
\end{pmatrix}}}

 \def\bea  {\begin{eqnarray}}   \def\eea  {\end{eqnarray}}

\newsavebox\MBox

\title{Cross-correlations as a Diagnostic Tool for Primordial Gravitational Waves }

\author[a]{Ameek Malhotra}\author[a]{, Ema Dimastrogiovanni}
\author[b,c]{, Matteo Fasiello}
\author[d]{and Maresuke Shiraishi}
\affiliation[a]{Sydney Consortium for Particle Physics and Cosmology\\ School of Physics, The  University of New South Wales, Sydney NSW 2052, Australia}
\affiliation[b]{
Instituto de F\'{i}sica T\'{e}orica UAM/CSIC, calle Nicol\'{a}s Cabrera 13-15, Cantoblanco, 28049, Madrid, Spain}
\affiliation[c]{Institute of Cosmology \& Gravitation, University of Portsmouth, PO1 3FX, UK}
\affiliation[d]{Department of General Education, National Institute of Technology, Kagawa College,
355 Chokushi-cho, Takamatsu, Kagawa 761-8058, Japan}

\emailAdd{ameek.malhotra@unsw.edu.au}
\emailAdd{e.dimastrogiovanni@unsw.edu.au}
\emailAdd{matteo.fasiello@csic.es}
\emailAdd{shiraishi-m@t.kagawa-nct.ac.jp}

\abstract{We explore and corroborate, by working out explicit examples, the effectiveness of cross-correlating stochastic gravitational wave background anisotropies with CMB temperature fluctuations as a way to establish the primordial nature of a given gravitational wave signal.~We consider the case of gravitational wave anisotropies induced by scalar-tensor-tensor primordial non-Gaussianity. Our analysis spans anisotropies exhibiting different angular behaviours, including a quadrupolar dependence.~We calculate the expected uncertainty on the non-linearity parameter $F_{\rm NL}$ obtained as a result of cross-correlation measurements for several proposed experiments such as the ground-based Einstein Telescope, Cosmic Explorer, and the space-based Big-Bang Observer.~As a benchmark for future survey planning, we also calculate the theoretical, cosmic-variance-limited, error on the non-linearity parameter.}

\begin{document}

\maketitle

\section{Introduction}
The ever-increasing number of detections of gravitational wave events generated by astrophysical sources has ushered in a new era for gravitational wave astronomy. The increased sensitivity of upcoming gravitational waves (GW) detectors  \cite{AmaroSeoane:2012km,Maggiore:2019uih,Reitze:2019iox,Sato:2017dkf} holds the potential to transform also our current picture of cosmology.
There are indeed several mechanism of GW  production in the early universe, from pre-heating to first order phase transition (see e.g. \cite{Caprini:2015tfa} for a compact review on the subject). Perhaps most notably, the existence of a stochastic primordial GW background is a universal prediction of inflation. 

The minimal scenario, single-field slow-roll inflation, is characterised by a slightly red-tilted gravitational wave spectrum. Such signal is out of reach for upcoming probes, with the possible exception, depending on the value of the tensor to scalar ratio ``$r$'', of the proposed Big Bang Observer \cite{Crowder:2005nr}. In this work we shall instead be concerned with the primordial GW spectrum generated by multi-field inflationary configurations. These are not just possible in light of observations, but also likely from the top-down perspective \cite{Baumann:2014nda}. Several such models support a GW spectrum that can be sufficiently large at high frequencies to warrant detection by laser interferometers such as LISA and the Einstein Telescope (ET). In this context the study of the amplitude, scale-dependence, polarisation, and higher-point functions (see e.g \cite{Adshead:2009bz,Bartolo:2018qqn}) of the GW signal is paramount for the characterisation of the different inflationary mechanisms. The cross-correlation of observables from independent probes of primordial physics represents a further step  towards identifying the origin of the GW signal.

The starting point of our analysis are intrinsic\footnote{To avoid any confusion, we clarify here the terminology used when referring to different types of anisotropies. By \textit{induced} anisotropies we shall mean those that affect the GW power spectrum as a result of GW modes propagating through structure (see \cite{Bartolo:2019yeu} for a thorough recent study of these effects). We term \textit{intrinsic} anisotropies those due to primordial mode couplings, whose contribution exists irrespective of propagation effects. Finally, \textit{statistical} anisotropies are those resulting, for example, from the existence of a preferred direction in the sky. Naturally, these properties are not mutually exclusive: in \textit{Section}~\ref{stat} we will present the analysis of statistical anisotropies that may well be intrinsic in origin.} anisotropies induced on the primordial GW spectrum by long-short mode couplings. The anisotropies in question arise from the existence of a non-trivial primordial correlator of the type $\langle \mathcal{O}_{k_1} \gamma_{k_2} \gamma_{k_3} \rangle $, where ``$\mathcal{O}$'' stands for a generic primordial fluctuation. It follows that they are a very useful  handle on inflationary interactions. As it turns out, they are particularly relevant at small scales. Whenever the $k$ modes in question are sufficiently large to have undergone sub-horizon evolution during e.g. radiation or matter domination, the non-zero primordial correlation is ``washed out" as a result of propagation effects on short modes, i.e. as a result of their different history \cite{Bartolo:2018evs}. It is therefore rather difficult to probe primordial non-Gaussianities at small scales. Nevertheless, this same line of reasoning suggests an intriguing exception: the \textit{ultra-squeezed} configuration. If one of the $k$ modes is sufficiently long (e.g. horizon-size) to avoid propagation effects, one can probe the primordial non-Gaussian signal by the anisotropy it induces on the GW power spectrum \cite{dimastrogiovanni_searching_2020}.  

It is important to stress that the squeezed configuration of non-Gaussianities is precisely that for which one ought to carefully identify the physical contribution to correlation functions. In light of so-called consistency relations (CRs), in place e.g. for scalar and scalar-tensor-tensor (STT) correlators in single field inflation, the leading contribution in the squeezed limit can be gauged away, leaving a sub-leading three-point function. Remarkably, CRs may be broken in a variety of scenarios, including those characterised by (i) multi-clock dynamics, (ii) excited initial states, (iii) alternative symmetry breaking patterns. It is these scenarios we shall have in mind throughout this work, both when studying sold inflation (a prototypical example of \textit{iii}) and when adopting a more phenomenological approach in \textit{Section}~\ref{stat}. 

Equipped with inflationary models supporting non-Gaussianities that induce anisotropies on the GW spectrum,  it behoves us to    
identify tools to help distinguish the signatures associated to different primordial set-ups as well to those of astrophysical origin. In addition to cosmological GW backgrounds, astrophysical backgrounds (AGWB) arising from unresolved sources are also expected to contribute to the stochastic GW background (SGWB) and its anisotropies \cite{Regimbau:2011rp,Cusin:2017fwz,Cusin:2017mjm,Pitrou:2019rjz,Bertacca:2019fnt}.  While the astrophysical signal is very interesting in its own right, it is expected to act as noise for other backgrounds making the latter difficult to identify. Extracting the underlying primordial signal thus requires a complete characterisation of the astrophysical background so that it can be ``subtracted'' away from the data. This might be possible with LISA, ET and CE (see  \cite{Regimbau:2016ike,Pan:2019uyn,Sharma:2020btq,Pieroni:2020rob,Biscoveanu:2020gds,Boileau:2020rpg,Martinovic:2020hru}). However, anisotropies of the cosmological SGWB will remain out of reach of all but the most high sensitivity and high resolution detectors \cite{Baker:2019ync,Alonso:2020rar}.

One intriguing possibility \cite{adshead_multimessenger_2020} relies on the cross-correlation\footnote{Using cross-correlations to extract weak signals from noise dominated maps has an illustrious history. In the case of the CMB, cross-correlations with large scale structure observables have been used to detect the secondary CMB anisotropies which are generated at low redshifts, the works in \cite{Crittenden:1995ak,Peiris_2000,Afshordi:2003xu,Afshordi:2004kz,Padmanabhan:2004fy,Ho:2008bz} being notable examples. As for astrophysical sources, since the energy density of the AGWB depends on the properties of the host galaxies and the distribution of large scale structure, it is expected to have cross-correlations with observables like galaxy number counts and weak lensing \cite{Cusin:2018rsq,Canas-Herrera:2019npr,Alonso:2020mva,Yang:2020usq}. For ground-based interferometers, PTAs as well as the proposed BBO/DECIGO, the SGWB itself is detected through cross-correlations among different detectors \cite{Maggiore:1999vm,Allen:1996vm,Hellings:1983,romano_detection_2017}.} of SGWB anisotropies, $\delta_{GW}$, with CMB temperature  anisotropies\footnote{See  \cite{Geller:2018mwu} for the use of cross-correlations in the context of early universe phase transitions.}, $\delta_{T}$. The leading contribution to CMB temperature anisotropies originates from scalar fluctuations ``$\zeta$'' during inflation. Whenever a non-trivial squeezed STT correlator (i.e. $\langle \zeta\gamma \gamma \rangle)$ is in place, it contributes to GW anisotropies $\delta_{GW}$. Both these observables are then sensitive to primordial scalar fluctuations. Their cross-correlation will help single out the primordial component in gravitational wave anisotropies. Given the projected sensitivities of laser interferometers, cross-correlation with additional probes of primordial physics may well be the most efficient way to identify GW anisotropies of cosmological origin. \\
 \indent Building in part on the results of \cite{adshead_multimessenger_2020}, in this work we explore cross-correlations from a number of different angles. First, we focus on a specific inflationary model, solid inflation \cite{endlich_solid_2013}, whose STT bispectrum can be large and exhibits a quadrupolar angular dependence. We estimate how well a cross-correlation measurement can constrain the  non-linearity parameter $\fnl$. We shall also adopt a more general, phenomenological, approach to the case of statistically anisotropic cross-correlations, i.e. those with non-vanishing off-diagonal $C_{\ell}$'s. The results for the corresponding uncertainty on non-Gaussianity are discussed at length. \\ Our findings on the uncertainty on the non-linearity parameter $\fnl$ have benefitted from the very recent work in \cite{Alonso:2020rar}, where the instrumental noise in interferometer networks was analysed.~In order to provide a complete picture and as a useful indicator for future survey planning, we also calculate the theoretical (i.e. cosmic-variance-limited (CVL-level)) error on $\fnl$. \\

\noindent This paper is organised as follows: in \textit{Section} \ref{sec2} we provide a brief introduction to (intrinsic) anisotropies and to the basics of the more general Boltzmann approach to SGWB anisotropies; in \textit{Section}~\ref{sec3} we review the solid inflation model; in \textit{Section}~\ref{sec:CrossCorr} we derive the cross-correlation between CMB and SGWB anisotropies supported by solid inflation; in \textit{Section}~\ref{sec5} the  uncertainty on the non-linearity parameter $\fnl$ that can achieved through cross-correlations is obtained. We provide estimates that include instrumental noise alongside those limited only by cosmic variance. \textit{Section}~\ref{stat} is devoted to a phenomenological take on the case of \textit{statistically} anisotropic models, for which we evaluate the constraining power of cross-correlation measurements on non-Gaussianities; in \textit{Section}~\ref{sec7} 
we discuss our findings, draw the conclusions and comment on future work; \textit{Appendix} A collects results on the parameter space of solid inflation that are relevant to our analysis; detailed result on (auto and) cross-correlations can be found in \textit{Appendix} B;  \textit{Appendix} C contains useful details on the calculation of the noise angular power spectra.

\section{SGWB Anisotropies}
\label{sec2}
Stochastic backgrounds of gravitational waves are typically characterized in terms of the normalised energy density per logarithmic interval of frequency  \cite{Maggiore:1999vm},
\begin{align}
\Omegagw(f) \equiv \frac{1}{\rho_{\text{\tiny cr}}}\frac{d \rhogw}{d \ln f} 
\end{align}
where $\rho_{\text{\tiny cr}}$ is the critical energy density and $\rhogw$ is the energy density of GW. To characterize the anisotropies, we allow for a direction dependence through $\hn$
\begin{align}
\label{eq:omegadef}
\Omegagw(f) \equiv \frac{1}{4\pi}\int d^2\hn \,\overline{\omega}_{\rm GW}(f)\Big[1+\delta_{\rm GW}(f,\hat{n})  \Big]\, \equiv \overline{\Omega}_{\rm GW}(f)\Big[1+\frac{1}{4\pi}\int d^2\hat{n}\, \delta_{\rm GW}(f,\hat{n})    \Big]
\end{align}  
where $\delta_{\rm GW}$ is the SGWB anisotropy and $\overline{\Omega}_{\text{\tiny GW}}(f) $ the isotropic component of the energy density.

\subsection{Anisotropies from primordial non-Gaussianity}
The anisotropies of the cosmological SGWB can be broadly classified into primordial (\textsl{intrinsic}) and \textsl{induced} by propagation. Intrinsic anisotropies, as the name suggests, are imprinted upon the GW at the time of production and their properties depend upon the production mechanism considered. On the other hand, induced anisotropies arise from the propagation of GW in an inhomogeneous universe and have a typical magnitude $\delta^{\rm ind}_{\text{\tiny GW}}\sim \sqrt{A_S}$ \cite{Alba:2015cms,Bartolo:2019oiq,Bartolo:2019yeu,DallArmi:2020dar,Domcke:2020xmn}, where $A_S$ is the amplitude of the scalar power spectrum. As we shall see next, if the primordial bispectrum in the squeezed limit is sufficiently large, the intrinsic contribution, $\delta^{\rm prim}_{\text{\tiny GW}}\sim F_{\rm NL}\sqrt{A_S}$, can dominate over induced anisotropies.

The effects of large primordial non-Gaussianities on the tensor power spectra have been previously considered in  \cite{Ricciardone:2017kre,dimastrogiovanni_searching_2020},  where the 3-pt function $\langle{\gamma_{\vq}\gamma_{\vk}\gamma_{-\vk-\vq}}\rangle$ in the squeezed limit ($q\ll k $) generates a local quadrupolar anisotropy in the tensor power spectrum due to the coupling between the long mode $q$ and the short modes $k$ \footnote{Anisotropies can also arise, due to a local $\langle \zeta^{3}\rangle$ bispectrum, in the energy density of GW generated at second order in perturbation theory from the curvature perturbation $\zeta$ \cite{Bartolo:2019zvb}.}. More recently, ref. \cite{adshead_multimessenger_2020} focused on the case of SGWB anisotropies arising from a primordial STT bispectrum $\langle{\zeta_{\vq}\gamma_{\vk}\gamma_{-\vk-\vq}}\rangle$. Following the notation in \cite{adshead_multimessenger_2020}, we briefly review here how the primordial squeezed STT bispectrum can lead to anisotropies in the GW energy density. To begin with, we denote the scalar and tensor power spectra from inflation as
\begin{align}
\langle{\zeta_{\vk}\,\zeta_{\vk}}\rangle' \equiv P_{\zeta}(k)\,,\quad\langle{\gamma_{\vk}\,\gamma_{\vk}}\rangle' \equiv P_{\gamma}(k)\,,
\end{align}
and the squeezed three-point function as
\begin{align}
\langle\zeta_{\vq\to 0}\,\gamma_{\vk_1}\,\gamma_{\vk_2}\rangle' \equiv B_{STT}(\vq,\vk_1,\vk_2)\,,
\end{align}
where the prime denotes the fact that these correlators are written without the $(2\pi)^3$ factor and the delta function  enforcing momentum conservation. The long-wavelength curvature fluctuation $\zeta$ modulates the  primordial tensor power spectrum as \cite{Jeong:2012df,Dai:2013kra,Brahma:2013rua,Dimastrogiovanni:2014ina,Dimastrogiovanni:2015pla}
\begin{align}
\label{eq:Pmod}
P^{\rm mod}_{\gamma}(\vk,\vx) = P_{\gamma}(k)\left[1+\int_{q \ll k}\frac{d^3q}{(2\pi)^3}\,e^{i\vq\cdot\vx}\fnl(\vq,\vk)\zeta(\vec{q})\right]\,,
\end{align}
where the $\vec{x}$-dependence indicates that the power spectrum is evaluated locally near $\vec{x}$, in a volume with linear dimensions smaller than $1/q_{}$. The parameter $\fnl$ is defined as 
\begin{align}
\label{eq:fnldef}
\fnl(\vq,\vk)\equiv \frac{B_{\rm STT}(\vq,\vk-\vq/2,-\vk-\vq/2)}{P_{\zeta}(q)P_{\gamma}(k)}\,.
\end{align}
The GW energy density observed today in terms of the primordial power spectrum is
\begin{align}
\Omegagw(k,\eta_0) = \frac{k^2}{12 a_0^2 H_0^2}\mathcal{T}^2(k,\eta_0)\cdot\frac{1}{4\pi}\int d^2\hn\, \mathcal{P}^{\rm mod}_{\gamma}(\vk,\vec{d},\eta_{\rm in})\,,
\end{align}
where $\mathcal{P}^{}_{\gamma}=({k^3}/2\pi^2)P^{}_{\gamma}$ and $\mathcal{T}(k,\eta_0)$ is the tensor transfer function that relates the mode $k$ at the time of horizon entry $\eta_{\rm in}$ to its present value. Note that the power spectrum is evaluated at the point $\vec{d}=-\hat{n}d$, where $d=\eta_0-\eta_{\rm in}$ is the time elapsed from horizon re-entry to the present and $\hat{n}$ is the direction of propagation of the GW mode ($\vec{k}=k\hat{n}$). Separating the isotropic and anisotropic contributions we obtain,
\begin{align}
\overline{\Omega}_{\text{\tiny GW}}(k,\eta_0) = \frac{k^2}{12 a_0^2 H_0^2}\mathcal{T}^2(k,\eta_0)\mathcal{P}_{\gamma}(k)
\end{align}
and
\begin{align}\label{cite1}
\delta_{\text{\tiny GW}}^{\rm prim}(k,\hat{n}) = \int_{q_{}\ll k}\frac{d^3q}{(2\pi)^3}\,e^{-id\hat{n}\cdot\vq}\fnl(\vq,\vk)\zeta(\vec{q})\,.
\end{align}
The presence of inhomogeneities in the primordial power spectrum present at the time of horizon re-entry leads to an anisotropy in the energy density associated with incoming GW modes. 

\subsection{Boltzmann approach to SGWB anisotropies}
\label{Boltz}
The Boltzmann formalism, typically used for the computation of the CMB anisotropies, has been recently applied to study the statistics of SGWB anisotropies  \cite{Contaldi:2016koz,Bartolo:2019oiq,Bartolo:2019yeu,DallArmi:2020dar}. Since this approach is particularly well suited for studying anisotropies of relic backgrounds, we show here how the $\delta_{\text{\tiny GW}}^{\rm prim}$ considered in the previous section fits into this picture. \\
\noindent Working in the Poisson gauge for the large scale perturbations one has
\begin{align}
ds^2 = a^2(\eta)[-e^{2\Phi}d\eta^2 + (e^{-2\Psi}\delta_{ij}+\chi_{ij})dx^i dx^j]\,,
\end{align}
where $\eta$ is the conformal time, $\Psi$ and $\Phi$ are the scalar potentials, and $\chi_{ij}$ are the transverse traceless tensor perturbations. We adopt here the geometrical optics limit \cite{PhysRev.166.1263,PhysRev.166.1272}. The propagation of gravitons can be approximately described as a stream of massless, collisionless particles that propagate along the null geodesics of the background and carry the energy flux of the GW. The distribution function of gravitons $f=f(x^{\mu},p^{\mu})$ can be defined as a function of their position $x^{\mu}$ and momentum $p^{\mu}=dx^{\mu}/d\lambda$, $\lambda$ being an affine parameter along the trajectory. It obeys the Boltzmann equation,
\begin{align}
	\mathcal{L}[f(\lambda)] = \mathcal{C}[f(\lambda)] +\mathcal{I}[f(\lambda)]\,,
\end{align}
with $\mathcal{L}=d/d\lambda$ the Liouville term and $\mathcal{C},\mathcal{I}$ the collision and emission terms respectively. In the absence of graviton collisions and treating the emission term as an initial condition on the distribution function, one arrives at the following equation \cite{Contaldi:2016koz,Bartolo:2019oiq,Bartolo:2019yeu}
\begin{align}
\frac{df}{d\eta} = \underbrace{\frac{\partial f}{\partial \eta}+\frac{\partial f}{\partial x^i}\frac{dx^i}{d\eta}}_{\text{free streaming}}\quad+\underbrace{\frac{\partial f}{\partial k}\frac{dk}{d\eta}}_{\text{SW, ISW and RS}}+\quad\underbrace{\frac{\partial f}{\partial n^i}\frac{dn^i}{d\eta}}_{\text{lensing}} = 0\,,
\end{align}
where $k=|p|a$ is the comoving momentum and $\hn=\hat{p}$ is the direction of motion. The distribution function can be expanded as, 
\begin{align} 
f(\eta,x^i,k,n^i) = \bar{f}(k) - k\frac{\partial \bar{f}}{\partial k}\Gamma(\eta,x^i,k,n^i)+\dots\,,
\end{align}
where $\bar{f}$ is the homogeneous and isotropic component which solves the zeroth order equation 
\begin{align}
\frac{\partial \bar{f}}{\partial\eta} = 0
\end{align}
and $\Gamma(\eta,x^i,k,n^i)$ parametrises the first order correction. The zeroth order equation is solved by any distribution that is a function of solely the comoving momentum $k$. The GW energy density observed today at a point $\vx_0$ can be easily calculated from the distribution function,
\begin{align}
\rhogw(\eta_0,\vx_0) = \frac{1}{a_0^4}\int d^3k\, k\, f(\eta_0,\vx_0,k,\hn) \equiv \rho_{\rm cr} \int d \ln k\,\Omegagw(k,\vx_0)\,.
\end{align}
From this one can define, 
\begin{align}
\delta_{\text{\tiny GW}} \equiv \frac{\omegagw(k,\hn,\vx_0)-\overline{\omega}_{\text{\tiny GW}}(k)}{\overline{\omega}_{\text{\tiny GW}}(k)}  = \left[4 - \frac{\partial \ln \overline{\Omega}_{\text{\tiny GW}}(k)}{\partial \ln k} \right]\Gamma(\eta_0,\vx_0,k,\hn)\,,
\end{align}
where $\omega_{\text{\tiny GW}}$ is defined as 
\begin{align}
	\Omega_{\text{\tiny GW}}(k,\vec{x}_0) = \frac{1}{4\pi}\int d^2\hn \,\omega_{\text{\tiny GW}}(k,\hn,\vec{x}_0)\,.
\end{align}
In the Fourier basis ($\vx_0\to \vq$), the first order Boltzmann equation is 
\begin{align}
\Gamma' + iq\mu\Gamma = S(\eta,\vq,\hn)\,,
\end{align}
where prime denotes a derivative w.r.t the conformal time $\eta$ and $\mu = \hq\cdot\hn$ is the projection of the wavevector $\vq$ on the line of sight direction $\hn$. The source term is
\begin{align}
	S = \Psi' -iq\mu\Phi-n^i n^j\chi_{ij}'/2\,.
\end{align} 
The solution for $\Gamma$ consists of three contributions: an initial condition term $\Gamma_I$, a scalar source term $\Gamma_S$, and a tensor term $\Gamma_T$
\begin{align}
\Gamma(\eta,\vq,k,\hn) = \Gamma_I(\eta,\vq,k,\hn) + \Gamma_S(\eta,\vq,\hn) + \Gamma_T(\eta,\vq,\hn) 
\label{eq:gamma3}\,.
\end{align}
The scalar and tensor terms correspond to the induced anisotropies arising from the propagation of GWs in a background with large-scale perturbations, much like the case of the CMB.  As for the initial condition, this term solves the homogeneous part of the first order Boltzmann equation,
\begin{align}
	\Gamma'+ iq\mu\Gamma=0\,,
\end{align}
such that
\begin{align}
\Gamma_I(\eta,\vq,k,\hn) = e^{iq\mu(\eta_{\rm in} - \eta)}\Gamma_I(\eta_{\rm in},\vq,k,\hn)\,.
\end{align}
The results presented so far in Sec.~\ref{Boltz} have been first derived in \cite{Contaldi:2016koz,Bartolo:2019oiq,Bartolo:2019yeu}. We can relate them to the primordial anisotropies discussed in the previous section. For anisotropies originating from squeezed STT non-Gaussianity, from Eq.~(\ref{cite1}) one finds
\begin{align}
\left[4 - \frac{\partial \ln \overline{\Omega}_{\text{\tiny GW}}(k)}{\partial \ln k} \right]\Gamma_{I}(\eta_{\rm in},\vq,k,\hn)=\fnl(\vq_{},\vk)\zeta(\vec{q}_{})
\label{eq:deltagamma}\,.
\end{align}
Note that for CMB photons the anisotropies are frequency independent at first order due to thermal initial conditions. This is not, however, the case for gravitons: a frequency dependence at first order may arise due to specific initial conditions (such as those from inflation, as we shall see in \textit{Section}~\hyperlink{section.3}{3}).~This notion is encoded \cite{Bartolo:2019yeu} in the frequency-dependence of the initial conditions term.

\section{Solid Inflation} 
\label{sec3}

For an inflationary model to be testable via CMB-SGWB cross-correlations as in \cite{adshead_multimessenger_2020}, it is essential for it to exhibit two main features : (i) a non-trivial STT squeezed bispectrum\footnote{By non-trivial we specifically refer to the part of the bispectrum that violates the single-field consistency relations \cite{maldacena_non-gaussian_2003,creminelli_single-field_2004}. When consistency relations are preserved, the leading-order contribution to the squeezed bispectrum amounts to a gauge artifact and its physical contribution to the anisotropies is suppressed. Besides solid inflation, CRs violation is possible in models with extra (spinning) fields \cite{Bordin:2018pca,Dimastrogiovanni:2018gkl,Iacconi:2019vgc,Iacconi:2020yxn}, non-Bunch Davies initial states \cite{Holman:2007na,Agarwal:2012mq,Akama:2020jko} and  alternative symmetry breaking patterns \cite{Bartolo:2015qvr,Ricciardone:2016lym,Ricciardone:2017kre,Celoria:2020diz}, to name but a few.};  (ii) a blue-tilted tensor power spectrum so that the isotropic background can be comfortably detected with interferometers. One such model is solid inflation \cite{endlich_solid_2013,endlich_squeezed_2014}, where the acceleration is driven by three scalar fields with derivative couplings and certain symmetries that resemble the effective field theory for a homogenous and isotropic solid \cite{Dubovsky:2005xd}. In contrast to the minimal single field models where one relates the adiabatic perturbations to the Goldstone bosons of the spontaneously broken time translation symmetry, here the fields $\phi^I$ are allowed to have space-dependent but time-independent background values,
\begin{align}
\langle{\phi^I}\rangle = x^I\,,
\label{eq:solid_symm}
\end{align} 
which break translational symmetry along $x$ (comoving coordinates). Homogeneity and isotropy of the background is recovered by requiring the existence of internal shift and rotational symmetries,
\begin{align}
\phi^I\to \phi^I + a^I,\quad a^I = \text{const.}\,,
\end{align} 
\begin{align}
\phi^I\to O^I_J\phi^J,\quad O^I_J \in SO(3)\,,
\end{align}
so that the background configuration \eqref{eq:solid_symm} is invariant under combined spatial and internal translations + rotations. Lastly, the model is equipped with an approximate internal dilation symmetry which ensures that there is no breakdown of the effective field theory of the solid in the expanding background,
\begin{align}
\phi^I \to \lambda\phi^I\,.
\end{align}
This also ensures that $\rho+p\ll\rho$, hence this ``solid'' can drive a period of exponential expansion.

\subsection{STT Bispectrum}
The alternative symmetry breaking pattern of solid inflation is such that it violates the single-field consistency relations, resulting in a squeezed-limit three-point function \cite{endlich_squeezed_2014},
\begin{align}
\label{eq:STTsolid}
\langle{\zeta_{\vq\to 0}\gamma_{\vk}^{\lambda}\gamma_{-\vk}^{\lambda'}}\rangle' = \frac{16}{9}\frac{F_Y}{F}P_{\zeta}(q)P_{\gamma}(k)\log\left({\frac{k}{aH}}\right)\left(\epsilon_{ij}^{\lambda}\epsilon_{ij}^{\lambda'}-3\hq_i \epsilon_{ij}^{\lambda}\epsilon_{jk}^{\lambda'}\hq_k\right)\,,
\end{align}
where $F,F_Y$ are parameters of the solid inflation Lagrangian, and the polarization tensor is normalised as $\epsilon^{\lambda}_{ij}\epsilon^{\lambda'}_{ij} = 2\delta_{\lambda\lambda'}$. With the STT bispectrum in (\ref{eq:STTsolid}), the amplitude $\fnl$ defined in \eqref{eq:fnldef} exhibits quadrupolar angular dependence through the combination $\hq\cdot\hn$ (having defined $\vec{k}=k\hn$) as:
\begin{eqnarray}
F_{\rm NL}(\vq,\vk) =&& \frac{16}{9}\frac{F_Y}{F}\log\left({\frac{k}{aH}}\right)\frac{1}{2}\sum_{\lambda,\lambda'}\left(\epsilon_{ij}^{\lambda}\epsilon_{ij}^{\lambda'}-3\hq_i \epsilon_{ij}^{\lambda}\epsilon_{jk}^{\lambda'}\hq_k\right)\nonumber\\&&=\widetilde{F}_{\rm NL}(k)\left[\frac{1}{5}\sum_{M}Y_{2M}(\hq)Y^*_{2M}(\hn)\right],
\label{equ:fnl}
\end{eqnarray}
where 
\begin{align}
\widetilde{F}_{\rm NL}(k) = 8\pi\cdot\frac{16}{9}\frac{F_Y}{F}\log\left({\frac{k}{aH}}\right)\,.
\label{eq:lambdadef}
\end{align}
Since the power spectra and bispectra are evaluated at times when the modes $k,q$ have exited the horizon $(k\ll aH)$ \cite{endlich_solid_2013,endlich_squeezed_2014}, we see that $\widetilde{F}_{NL}<0$.~In order to arrive at (\ref{equ:fnl}), we made use of the relations $\sum_{\lambda,\lambda'}\left(\epsilon_{ij}^{\lambda}\epsilon_{ij}^{\lambda'}-3\hq_i \epsilon_{ij}^{\lambda}\epsilon_{jk}^{\lambda'}\hq_k\right)=2\cdot(3\cos^2{\theta}-1)$ and $\mathcal{P}_2(\hq\cdot\hn) = (4\pi/5)\sum_{M}Y_{2M}(\hq)Y^*_{2M}(\hn)$, where $\mathcal{P}_2$ is the second Legendre polynomial and the $Y_{LM}$'s are the spherical harmonics.\\

\noindent As is customary for such a modulation, and for the sake of comparison with other instances of quadrupolar anisotropies generated by primordial non-Gaussianity, we can expand it in spherical harmonics  
\begin{align}
P_{\gamma}(\vk,\vx) = P_{\gamma}(k)\left[1+\sum_{M}Q_{2M}(k,\vx)Y_{2M}(\hk)\right]\,,
\end{align}
where one finds
\begin{align}
Q_{2M}(k,\vx)= \frac{1}{5}\int\frac{d^{3}q}{(2\pi)^{3}}\,e^{i\vec{q}\cdot\vec{x}}\, \widetilde{F}_{\rm NL}(q,k)Y_{2M}^{*}(\hat{q})\zeta(\vec{q})  \,.
\end{align}

\subsection{Blue tensor spectrum}
Besides the breaking of consistency relations, another interesting feature of solid inflation is that the tensor power spectrum is blue tilted ($n_T\simeq 2\epsilon_p c_{L}^2$) \cite{endlich_solid_2013}. Its amplitude can be written as
\begin{align}
	\mathcal{P}_{\gamma}(k) = A_T\left(\frac{k}{k_{ p}}\right)^{n_T},\quad {\rm with} \quad	A_T\simeq \frac{H^2_{ p}}{\pi^2 M^2_{\rm Pl}}\left(\frac{\tau_{ p}}{\tau_{ e}}\right)^{8c_{T}^2\epsilon/3}.
\end{align}
The subscript `p' is a reminder that these quantities are evaluated at the conformal time when the pivot scale exits the horizon, $\tau_{ e}$ indicates the end of inflation and $M_{\rm Pl}^2 = (8\pi G)^{-1} $. The tranverse and longitudinal sound speeds $(c_{T},c_{L})$ are related to each other via $c_T^2 \simeq 3/4\left(1+c_L^2-2\epsilon/3\right)$, see \cite{endlich_solid_2013}.

For a GW signal to be detected by interferometers, the energy density $\Omegagw$ needs to be sufficiently large in the frequency range where the detectors have the highest sensitivity. This does not typically occur in the simplest single-field slow-roll inflationary models, which predict a red tilted power spectrum ($n_T=-r/8$) and, consequently, a suppression of power at small scales. In solid inflation the tensor power spectrum is blue tilted and, if sufficiently large, the resulting SGWB can be observable by DECIGO and BBO (see Fig.~\ref{fig:SolidGW}), this without violating any of the present CMB constraints (see Appendix \ref{appA} for more details). We should stress here that the next generation of CMB experiments e.g. LiteBIRD \cite{Matsumura:2013aja}, CMB-S4 \cite{Abazajian:2016yjj} will be sensitive to $r\gtrsim 0.001$ and thus will be able to detect or rule out the $r=0.07$ value considered here.

\subsection{The multi-field isotropic model}
In describing our results for solid inflation, we shall find it useful to compare it to a specific multi-field inflationary scenario that we will refer to as the ``isotropic model''. This nomenclature is there to underscore the contrast with the quadrupolar anisotropy  characterising the STT correlator of solid inflation. First introduced in \cite{Bordin:2018pca}, this inflationary scenario arises from an effective theory description of a massive spin-2 particle non-minimally coupled to the inflaton background. The direct (as opposed to the minimal, gravity-mediated, one) coupling effectively relaxes unitarity constraints on the mass of the spin-2 field and allows for an intriguing phenomenology.

 The multi-field set-up delivers a non-trivial STT bispectrum in the squeezed configuration, and is therefore is a natural candidate \cite{adshead_multimessenger_2020} for the study of mechanisms supporting intrinsic GW  anisotropies. We refer the interested reader to \cite{Bordin:2018pca} for a thorough description of the multi-field EFT approach and to \cite{adshead_multimessenger_2020} for its use in the context of cross-correlations. It suffices to say here that a tensor spectral index of $n_{T}\sim 0.27$ fits well within the viable \cite{Iacconi:2019vgc,Iacconi:2020yxn} parameter space of this set-up. This is the value of $n_T$ we shall be using throughout the paper for comparison with, for example, the solid inflation scenario.

\begin{figure}
	\centering
	\includegraphics[width=0.70\linewidth]{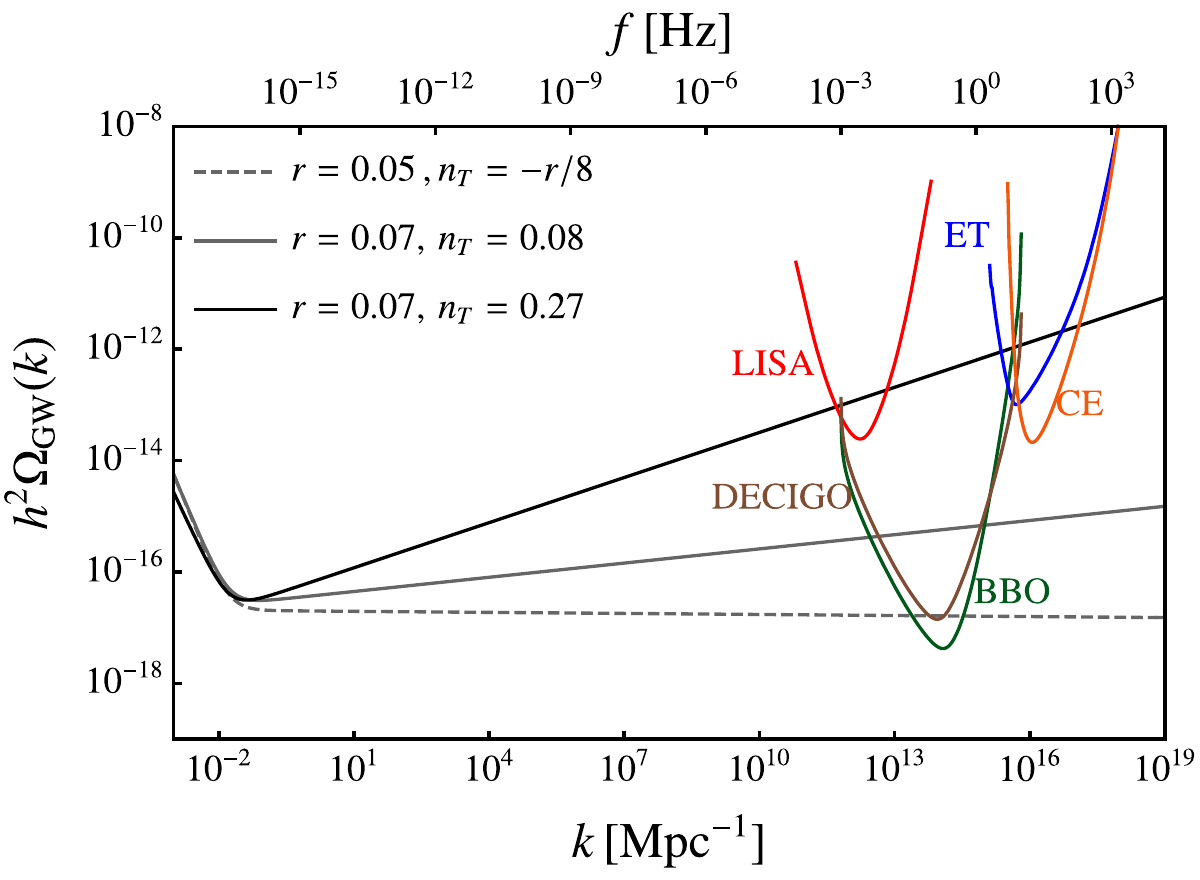}
	\caption[]{$\Omegagw(k)$ from solid inflation (gray curve), the ``isotropic'' model (black curve) and from single-field slow-roll inflation (dashed curve)  plotted alongside the power-law sensitivity curves for LISA, ET, CE, BBO and DECIGO\footnotemark. The tensor-to-scalar ratio $r$ is evaluated at the pivot scale $k_{\rm p}=0.05\kmpc$.}
	\label{fig:SolidGW}
\end{figure}
\footnotetext{For the sensitivity curves, we followed \cite{smith_lisa_2019,robert_caldwell_2019_3337373} for LISA and used the fits provided in  \cite{schmitz_new_2020,schmitz_kai_2020_3689582} for ET, CE, BBO and DECIGO.}

\section{Correlation between CMB and SGWB anisotropies}
\label{sec:CrossCorr}

We now compute the cross-correlation of anisotropies in the CMB temperature with those in the SGWB for the case of solid inflation (Section~\hyperlink{section.4.1}{4.1}). We also compare the magnitude of this cross-correlation with the cross-correlation of the CMB and the induced SGWB anisotropies (Section~\hyperlink{section.4.2}{4.2}).

\subsection{Primordial anisotropies}
We begin with the correlation between the CMB temperature anisotropy $\deltaT$ and the primordial SGWB anisotropiy $\delta^{\rm prim}_{\text{\tiny{GW}}}$. For ease of notation we drop the superscript \textsl{prim} and simply denote it as $\deltagw$
\begin{align}
\deltagw(k,\hx) = \int_{q\ll k}\frac{d^3q}{(2\pi)^3}e^{i\vq\cdot\vec{x}}\fnl(\vq,\vk)\zeta(\vq)\,,
\label{eq:delta_GW} 
\end{align}
where, throughout the rest of the paper, we define $\vec{x}\equiv-d\hat{n}$, with $d=\eta_0-\eta_{\rm in}$. The spherical harmonics coefficients are
\begin{align}
\deltagw_{\ell m} = \int d^2\hx\,Y^*_{\ell m}(\hx)\int_{q\ll k}\frac{d^3q}{(2\pi)^3}e^{i\vq\cdot \vx}\fnl(\vq,\vk)\zeta(\vq)\,.
\end{align}
Similarly, the coefficients of the temperature anisotropies in the Sachs-Wolfe limit are \cite{Dodelson:2003ft},
\begin{align}
\deltaT_{\ell m} = \frac{4\pi}{5}i^\ell\int \frac{d^3p}{(2\pi)^3}Y^{*}_{\ell m}(\hp)j_{\ell}(pr_{\rm lss})\zeta(\vp)\,,
\end{align}
with $r_{\rm lss}$ the comoving distance to the last scattering surface. Here we simply state the final result for the cross-correlation and leave the details of the calculation for Appendix \ref{appB1} We find
\begin{align}
\qop{\deltagw_{\ell_1 m_1}\delta^{*\text{\tiny T}}_{\ell_2 m_2}}&\equiv C_{\ell_1}^{\text{\tiny GW-T}}\delta_{\ell_1, \ell_2}\delta_{m_1 ,m_2}=\sum_{L}i^{L- \ell_1}h_{2 L \ell_1}^{2}\frac{G_{L \ell_1}}{2\ell_1+1} \delta_{\ell_1, \ell_2}\delta_{m_1 ,m_2} \,,
\label{eq:cross_result}
\end{align}
where the function $G_{\ell_1 \ell_2}$ is defined as
\begin{align}
G_{\ell_1 \ell_2} &=  \frac{2}{25\pi}\int_{q\ll k} q^2dq\,j_{\ell_1}(q d)j_{\ell_2}(qr_{\rm lss})\widetilde{F}_{\rm NL}(q,k)P_{\zeta}(q)
\label{eq:crossintegral}
\end{align}
and
\begin{align}
h_{\ell_1 \ell_2 \ell_3 } \equiv \sqrt{\frac{(2\ell_1+1)(2\ell_2+1)(2\ell_3+1)}{4\pi}}\tj{\ell_1}{\ell_2}{\ell_3}{0}{0}{0}\,.
\label{eq:hdef}
\end{align}
\begin{figure}
	\centering
	\includegraphics[width=0.60\linewidth]{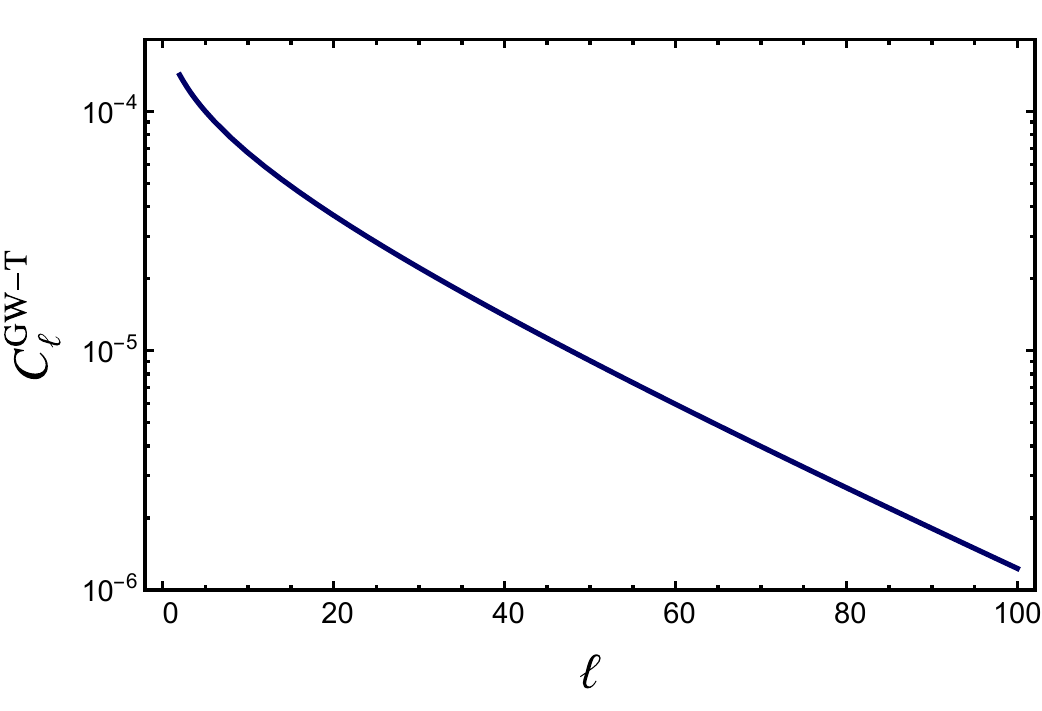}
	\caption{Magnitude of the cross-correlation of the primordial GW anisotropies from solid inflation with the CMB anisotropies, taking $|\widetilde{F}_{\rm NL}| =1=\,A_S$.}
	\label{fig:clcross}
\end{figure}	
We evaluate the integral in \eqref{eq:crossintegral} at interferometer scales with $k \sim 10^{13},\,10^{15},\,10^{17}$ Mpc$^{-1}$ which are the typical wavenumbers for LISA, DECIGO and ET respectively. We assume a scale invariant  scalar power spectrum $P_{\zeta}(q)=(2\pi^2/q^3)A_S$ (a good approximation for the solid inflation case \cite{endlich_solid_2013}). The cross-correlation in (\ref{eq:cross_result}) turns out to be positive for $\tilde{F}_{\text{NL}}<0$; Fig.~\ref{fig:clcross} provides the plot of $C_{\ell}^{\text{\tiny GW-T}}$ as a function of multipoles $\ell$, for $|\tilde{F}_{\text{NL}}|=A_{S}=1$. \\
\begin{figure}
	\centering
	\includegraphics[width=0.60\linewidth]{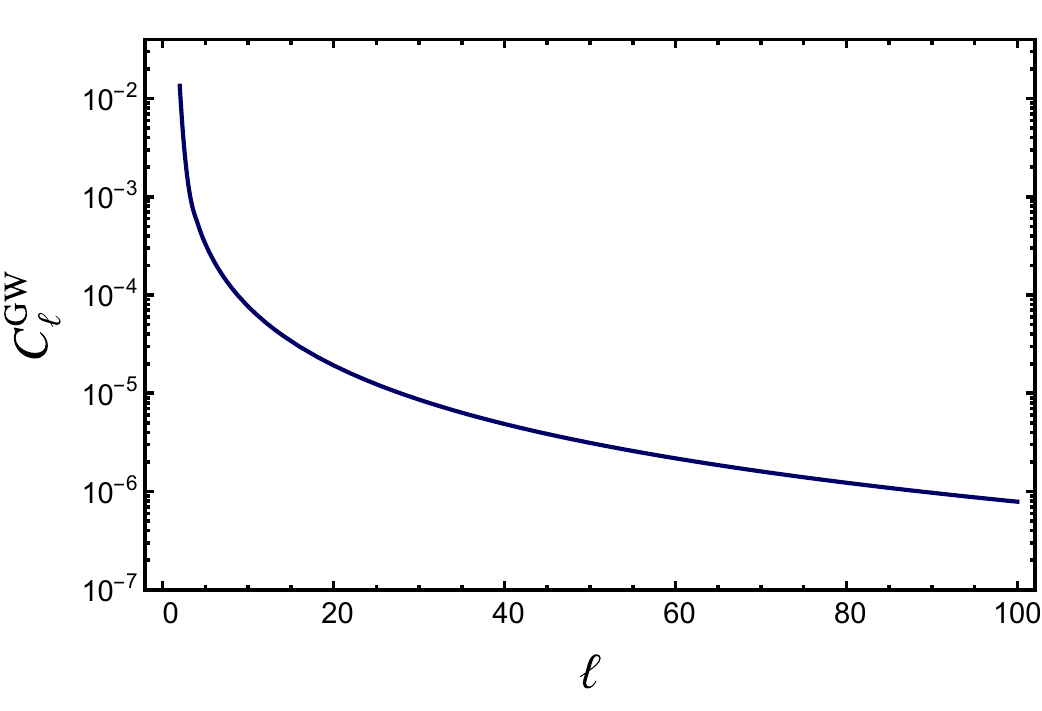}
	\caption{The auto-correlation of the primordial GW anisotropies from solid inflation taking $|\widetilde{F}_{\rm NL}| =1= A_S $.}
	\label{fig:clauto}
\end{figure}

\noindent In Figure \ref{fig:clauto} we plot the auto-correlation of the GW anisotropies,
\begin{align}
\qop{\deltagw_{\ell_1 m_1} \delta^{*\text{\tiny GW}}_{\ell_2 m_2}} \equiv C^{\text{\tiny GW}}_{\ell_1}\delta_{\ell_1,\ell_2}\delta_{m_1,m_2}\,,
\end{align}
where
\begin{equation}
C^{\text{\tiny GW}}_{\ell_1}=\sum_{L_1,L_2}i^{L_1-L_2}h_{\ell_1 L_1 2}^2 h_{{ \ell_1} L_2 2}^2\frac{H_{L_1 L_2}}{(2\ell_1+1)^2} 
\end{equation}
and
\begin{align}
H_{L_1 L_2} &\equiv  \frac{2}{25\pi}\int q^2dq\,j_{L_1}(q d)j_{L_2}(q {d} )\widetilde{F}^2_{\rm NL}(q,k)P_{\zeta}(q)\label{ref-e}
\end{align}
(we refer the reader to Appendix~\ref{appB2} for the derivation of these results). For $|\widetilde{F}_{\rm NL}| \gg 1$, one has $C_{\ell}^{\text{\tiny GW}} \gg A_S$ (roughly the value for the induced anisotropy, see \cite{Bartolo:2019oiq,Bartolo:2019yeu,DallArmi:2020dar,Domcke:2020xmn}), i.e. an intrinsic anisotropy that is much larger than the anisotropy induced by propagation.

\subsection{Induced anisotropies}
The dominant contribution to the induced anisotropies on large scales comes from the propagation of GW through the perturbed background. This corresponds to the term $\Gamma_{S}$ in \eqref{eq:gamma3} which, in the SW limit, is given by \cite{Bartolo:2019yeu,DallArmi:2020dar} 
\begin{align}
\Gamma_{\ell m, S} &= \int d^2\hx \,Y^{*}_{\ell m}(\hx)\int \frac{d^3q}{(2\pi)^3}e^{i\vq\cdot\vx}\cdot\frac{2}{3}\zeta({\vq})\nonumber\\
&=4\pi i^{\ell}\int \frac{d^3q}{(2\pi)^3}Y^*_{\ell m}(\hq)j_{\ell}(qd)\cdot\frac{2}{3}\zeta({\vq})\,.
\end{align}
The factor of 2/3 is due to the relation between $\zeta$ and $\Phi$ on superhorizon scales during the radiation dominated era. The corresponding SGWB anisotropy, $\deltagw_{\ell m} = (4-n_{\rm T})\Gamma_{\ell m, S}\simeq 4\Gamma_{\ell m, S}$, is also correlated with the CMB temperature anisotropies,
\begin{align}
\langle \deltagw_{\ell_1 m_1} \delta^{*\text{\tiny T}}_{\ell_2 m_2}\rangle = \frac{16}{15\pi}\delta_{\ell_1,\ell_2}\delta_{m_1, m_2}\int q^2dq\,j_{\ell_1}(qd)j_{\ell_2}(qr_{\rm lss})P_{\zeta}(q)\,.
\end{align} 
\begin{figure}
	\centering
	\includegraphics[width=0.60\linewidth]{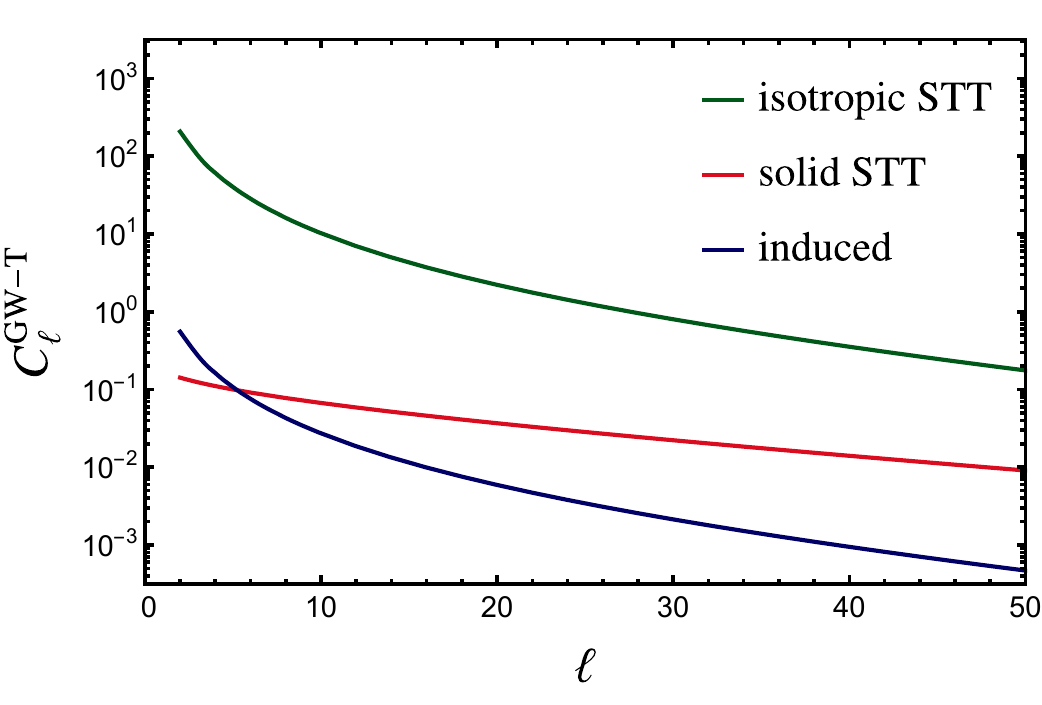}
	\caption{Plot of the induced cross-correlation along with the primordial ones for $\{\widetilde{F}_{\rm NL}=10^3,\,A_S = 1\}$. ``Isotropic STT'' refers to the cross-correlation derived in \cite{adshead_multimessenger_2020} (which does not exhibit angular dependence on $\hq,\hk$).}
	\label{fig:comparison}
	\end{figure}
\noindent Fig.~\ref{fig:comparison} shows a comparison of this cross-correlation (blue line) with the primordial one \eqref{eq:cross_result} (red line). For $|\widetilde{F}_{\rm NL}|=10^3$, the primordial contribution dominates except for the very first multipoles, while the induced contribution starts to dominate when $|\widetilde{F}_{\rm NL}|$ drops to order $10^2$. In the case of solid inflation, the non-trivial angular dependence of the STT bispectrum leads to a suppression in the cross-correlation compared to the bispectrum considered in \cite{adshead_multimessenger_2020}, which exhibited no angular dependence on $\hq,\hk$. For the latter kind of bispectrum, as long as $|\fnl|\gg 1$, the primordial contribution to the cross-correlation (green line of Fig.~\ref{fig:comparison}) dominates over (and it has the same $\ell$-dependence as) the induced one. It is important to point out that the $\ell$-dependence of the solid inflation cross-correlation  differs from that of both the isotropic and the induced contributions. As such, it should be seen as a useful distinctive feature associated to this model.

\section{\texorpdfstring{Projected constraints on the non-linearity parameter $\fnl$}{Error estimate for Fnl}}
\label{sec5}

\begin{figure} [h]
	\begin{minipage}{0.45\textwidth}
	\includegraphics[width=0.95\linewidth]{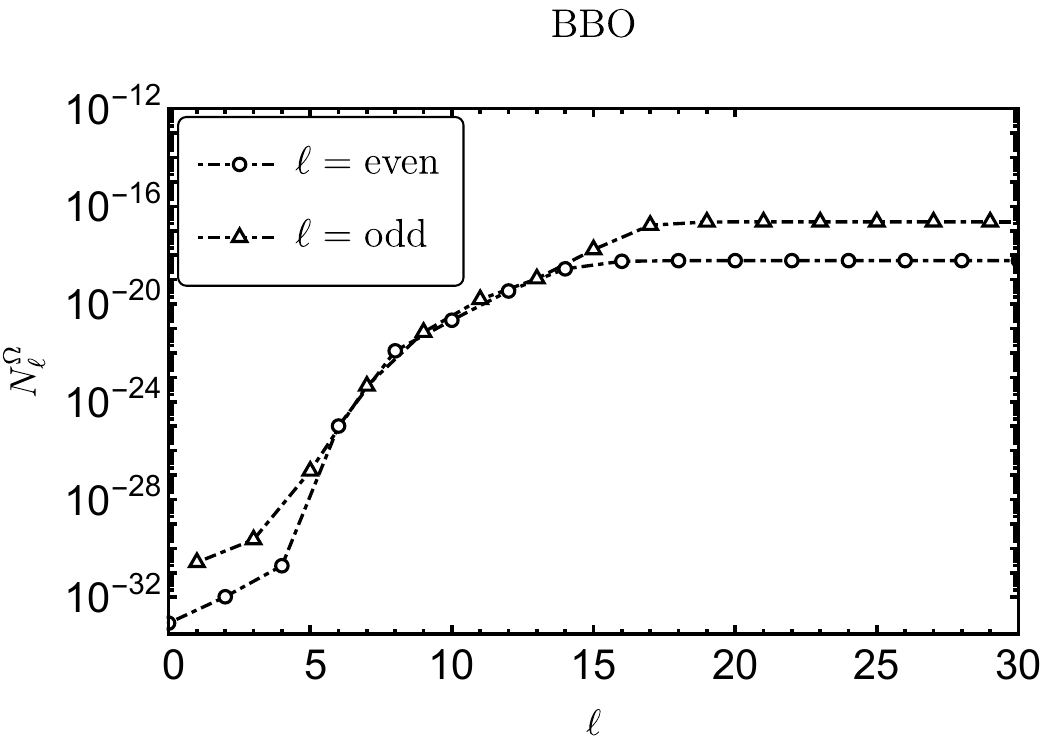}
	\end{minipage}
	\begin{minipage}{0.45\textwidth}
		\includegraphics[width=0.95\linewidth]{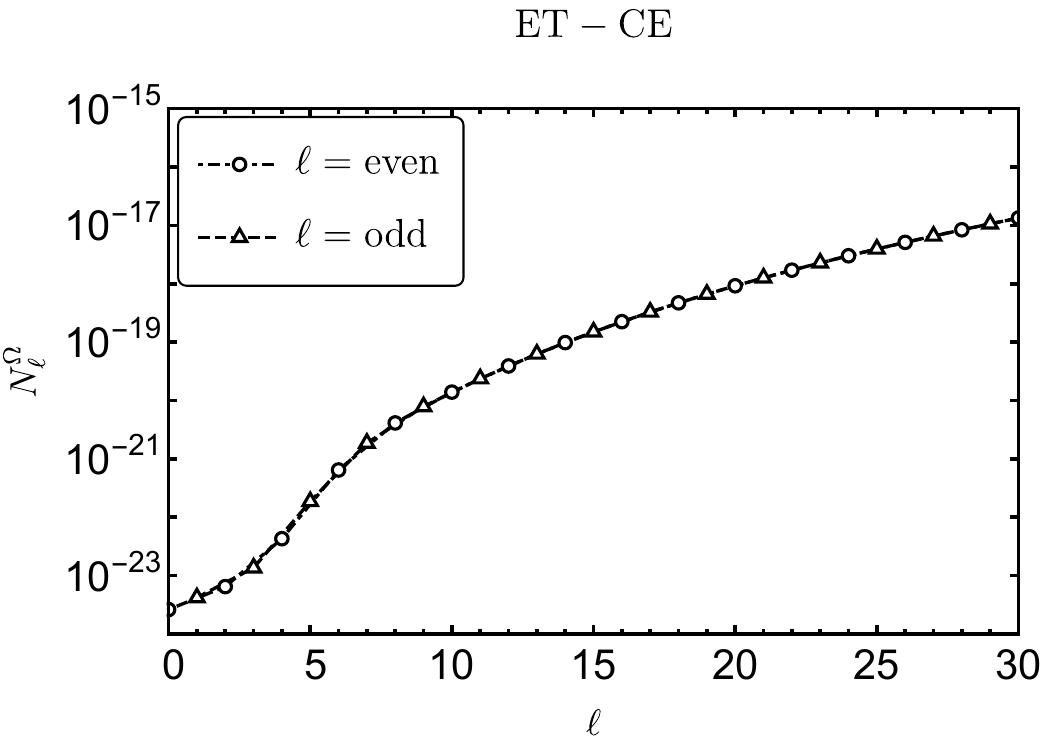}
	\end{minipage}
	\caption{The noise angular power spectra for the BBO star  configuration (left panel) and the ET-CE network (right). The quantity $N^{\rm GW}_{\ell}$ is defined via  $N^{\rm GW}_{\ell}= N^{\Omega}_{\ell}/\bar{\Omega}_{\rm GW}^2$.~Notice that the power spectra dramatically increases after the first few multiples.}
\label{Nlps}
\end{figure}

In this section we obtain the error in the determination of $\widetilde{F}_{\rm NL}$ through the cross-correlation $\GT$. The Fisher matrix is given by \cite{Verde_2010},
\begin{align}
F_{ ij} = \sum_{X Y}\sum_{{\ell =}\ell_{\rm min}}^{\ell_{\rm max}}\frac{\partial C^{X}_{\ell}}{\partial \theta_i}\left(\mathscr{C}^{XY}\right)^{-1} \frac{\partial C^{Y}_{\ell }}{\partial \theta_j}\,,
\end{align}
where $X,Y = \{{\rm TT,GW,GW-T}\}$ and $\theta_i$ is the set of parameters being measured. The matrix $\mathscr{C}$ reads,
\begin{align}
\mathscr{C}^{XY} = \frac{2}{2\ell+1}\begin{bmatrix}
(\TT)^2 & (\GT)^2 & \TT \GT \\
(\GT)^2 & (\GG)^2 & \GG\GT \\
\TT\GT & \GG\GT & \frac{1}{2}(\GT)^2 + \frac{1}{2}\TT\GG
\end{bmatrix}\,.
\end{align}
The Fisher matrix for the single parameter $\widetilde{F}_{\rm NL}$ is then
\begin{align}
F = \sum_{{\ell =} \ell_{\rm min}}^{\ell_{\rm max}} (2\ell+1)\frac{\left[(C_{\ell}^{\text{\tiny GW-T}})^2+C^{\text{\tiny TT}}_{\ell}C^{\text{\tiny GW}}_{\ell}\right]}{\left[(C_{\ell}^{\text{\tiny GW-T}})^2-C^{\text{\tiny TT}}_{\ell}C^{\text{\tiny GW}}_{\ell}\right]^2}I_{\ell}^2\,,
\label{eq:fullfisher}
\end{align}
where 
\begin{align}
I_{\ell} =  \frac{4\pi}{25}A_{\rm S}\sum_{L}i^{L-\ell}\frac{h_{2 L \ell}^2}{2\ell+1}\int \frac{dq}{q}j_{L}(qd)j_{\ell}(qr_{\rm lss})\quad \text{and } \TT = \frac{2\pi A_{\rm S}}{25\ell(\ell+1)}\,.
\end{align}
We estimate the error under the null hypotheses $\widetilde{F}_{\rm NL}=0$. For $(C_{\ell}^{\text{\tiny GW-T}})^2\ll  C^{\text{\tiny GW}}_{\ell} C^{\text{\tiny TT}}_{\ell} $ Eq.~\eqref{eq:fullfisher} can be simplified to
\begin{align}
F =\sum_{{ \ell =}\ell_{\rm min}}^{\ell_{\rm max}} (2\ell+1)\frac{I_{\ell}^2}{C^{\text{\tiny GW}}_{\ell} C^{\text{\tiny TT}}_{\ell}}\,.
\label{eq:error_def}
\end{align}
For GW anisotropies we first adopt the conservative assumption of a noise-dominated anisotropy map, i.e.  $\GG \simeq N_{\ell}^{\text{\tiny GW}}$. Next, in order to calculate
 $N_{\ell}^{\text{\tiny GW}}$ we follow  
  \cite{Alonso:2020rar} and use the associated code schNell\footnote{\url{https://github.com/damonge/schNell}}. The GW detectors we consider here are (i) a ground network consisting of ET and CE which will operate in the frequency range $1\operatorname{--}10^3$ Hz and (ii) BBO in the deci-Hz range  ($10^{-3}\operatorname{--}1$ Hz). DECIGO is similar in design and sensitivity to BBO and therefore we do not consider it separately. As for LISA, while it will be able to detect the monopole of the background in the $n_T=0.27$ case, its angular resolution turns out not to be sufficient for this kind of measurements unless used in conjunction with another detector in the mHz range \cite{Alonso:2020rar}. In Fig.~\ref{Nlps} we plot the noise angular power spectra for both experimental set-ups.
We refer the reader to Appendix \ref{appC} for details on the $N_{\ell}^{\text{\tiny GW}}$ calculations. 

\bigskip

\begin{figure}[h]
	\centering
	\includegraphics[width=\linewidth]{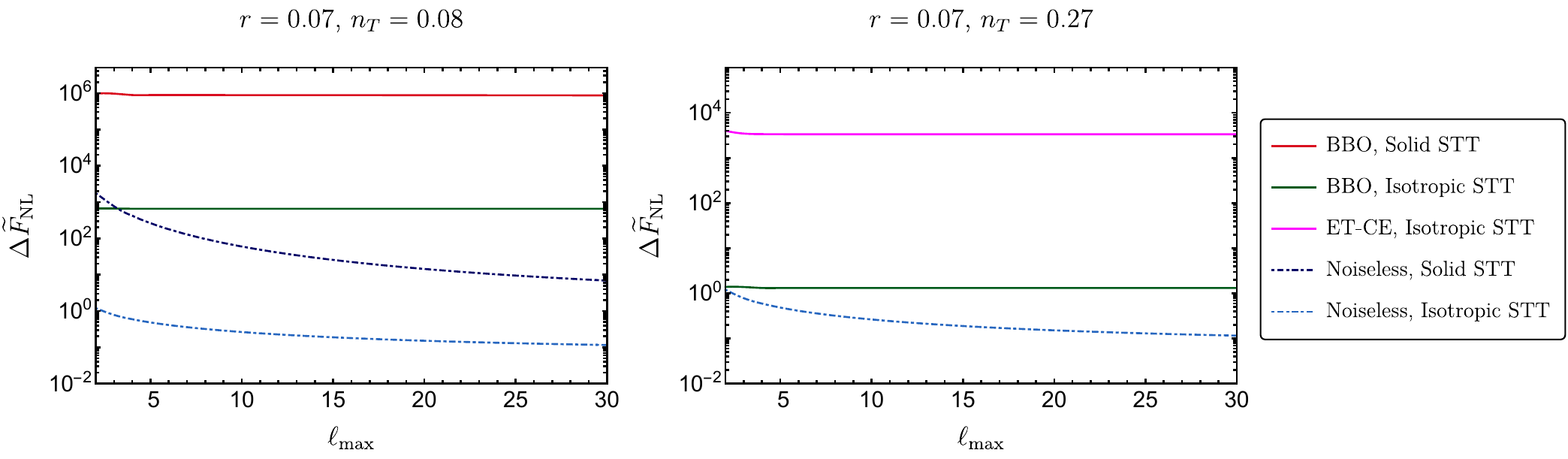}
	\caption{Expected $1\sigma$ error in measuring $\widetilde{F}_{\rm NL}$ as a function of $\ell_{\rm max}$ for BBO, ET-CE, and the idealized noiseless scenario. Both the isotropic and solid inflation cases are plotted for $n_T=0.08$ (left) whilst the isotropic model can support also $n_T=0.27$ (right panel).
	}
	\label{fig:errorfnl}
\end{figure}

We calculate the expected error in measuring ${F}_{\rm NL}$, i.e. $\Delta\widetilde{F}_{\rm NL} = (F)^{-1/2}$, using \eqref{eq:error_def} for $A_{\rm S}=2.1\times 10^{-9}$ and plot it in Figure~\ref{fig:errorfnl} as a function of $\ell_{\rm max}$. For the sake of comparison, in the same plot we provide the uncertainty also for the isotropic multi-field scenario. 
Given current constraints on solid inflation, the relative error on the non-linearity parameter $\tilde{F}_{\rm NL}$ is much larger compared to the isotropic case. This is due also to a considerably smaller $C_{\ell}^{\text{\tiny GW-T}}$ (see Figure~\ref{fig:comparison}). For the isotropic case, BBO will be able to accurately constrain $\widetilde{F}_{\rm NL}$. In particular, for e.g.
 $n_T=0.27, |\tilde{F}_{\rm NL}|\sim 10^3$, one obtains 
 \bea
 \frac{\Delta \tilde{F}_{\rm NL}}{\tilde{F}_{\rm NL}} \sim \mathcal{O}(0.001), 
  \eea
 while a relative error several orders of magnitude larger is found for solid inflation. The less sensitive ET-CE set-up  delivers a relative uncertainty of order one for the isotropic case.\\

Given that increasing the angular resolution of GW detectors could provide extremely tight constraints on $\widetilde{F}_{\rm NL}$, we find it worthwhile to work out the uncertainty in the idealized case of a noiseless GW survey. In such scenario the error is limited by cosmic variance alone. This configuration serves as a theoretical benchmark for what is achievable by cross-correlation measurements. For a noiseless survey we shall set $C^{\text{\tiny GW}}_{\ell} = C^{\text{\tiny GW,induced}}_{\ell}$ in the Fisher matrix calculations with 
\begin{align}
	C^{\text{\tiny GW,induced}}_{\ell}=\left(4-n_T\right)^2\frac{8\pi A_{\rm S}}{9\ell(\ell+1)}.
\end{align}
An analytical estimate of $\Delta\widetilde{F}_{\rm NL}$ can be obtained for the isotropic STT bispectrum if we adopt the approximation $d\simeq r_{\rm lss}$ in the expression for $I_{\ell}$. This quantity is given by \cite{adshead_multimessenger_2020}:
\begin{align}
	I_{\ell} =  \frac{4\pi}{5}A_{\rm S}\int \frac{dq}{q}j_{\ell}(qd)j_{\ell}(qr_{\rm lss})
	  \simeq \frac{4\pi}{5}\frac{A_{\rm S}}{2\ell(\ell+1)}.
\end{align} 
One obtains
\begin{align}
	F = \frac{9(\ell_{\rm max}-1)(\ell_{\rm max}+3)}{64},
\end{align}
and, for the error,
\begin{align}
	\Delta\widetilde{F}_{\rm NL} = F^{-1/2} \simeq \frac{8}{3 \ell_{\rm max}}.
\end{align} 
An ideal noiseless survey leads to a relative error of order $\mathcal{O}(10^{-4})$ for the isotropic scenario and  $\mathcal{O}(10^{-2})- \mathcal{O}(10^{-3})$ for solid inflation taking, as before, $|\tilde{F}_{\rm NL}|\sim 10^3$, and $\ell_{\rm max} \sim 30$ for the highest multiple (see Fig.~\ref{fig:errorfnl}).

\subsection{Cross-Correlations vs Auto-correlation measurement}

It is important to stress at this stage that the use of cross-correlations is particularly relevant (vis-a-vis auto-correlations) whenever  the GW anisotropy map is noise dominated, i.e. $N_{\ell}^{\text{\tiny{GW}}}\gtrsim C_{\ell}^{\text{\tiny{GW,prim}}}$. Indeed, in this regime autocorrelation measurements do not provide clear information on the magnitude of the anisotropies since the signal is  ``underneath'' the noise. In such a situation, cross-correlations of the GW and CMB anisotropies may be used to detect GW anisotropies of primordial origin. 

Whenever, instead, the signal dominates the noise i.e. $C_{\ell}^{\text{\tiny{GW,prim}}} \gg N_{\ell}^{\text{\tiny{GW}}} $, autocorrelations can be the go-to instrument to accurately constrain the primordial $\widetilde{F}_{\rm NL}$. As we illustrate below, for the cases  under scrutiny in this work, an adequate signal from auto-correlations corresponds to rather large values for $r,n_T$ and $\widetilde{F}_{\rm NL}$. This range of values, especially  the  $\widetilde{F}_{\rm NL}$ interval, is obtained only from scanning a very narrow region of the models parameter space.

\begin{figure}
	\centering
	\includegraphics[width=0.60\linewidth]{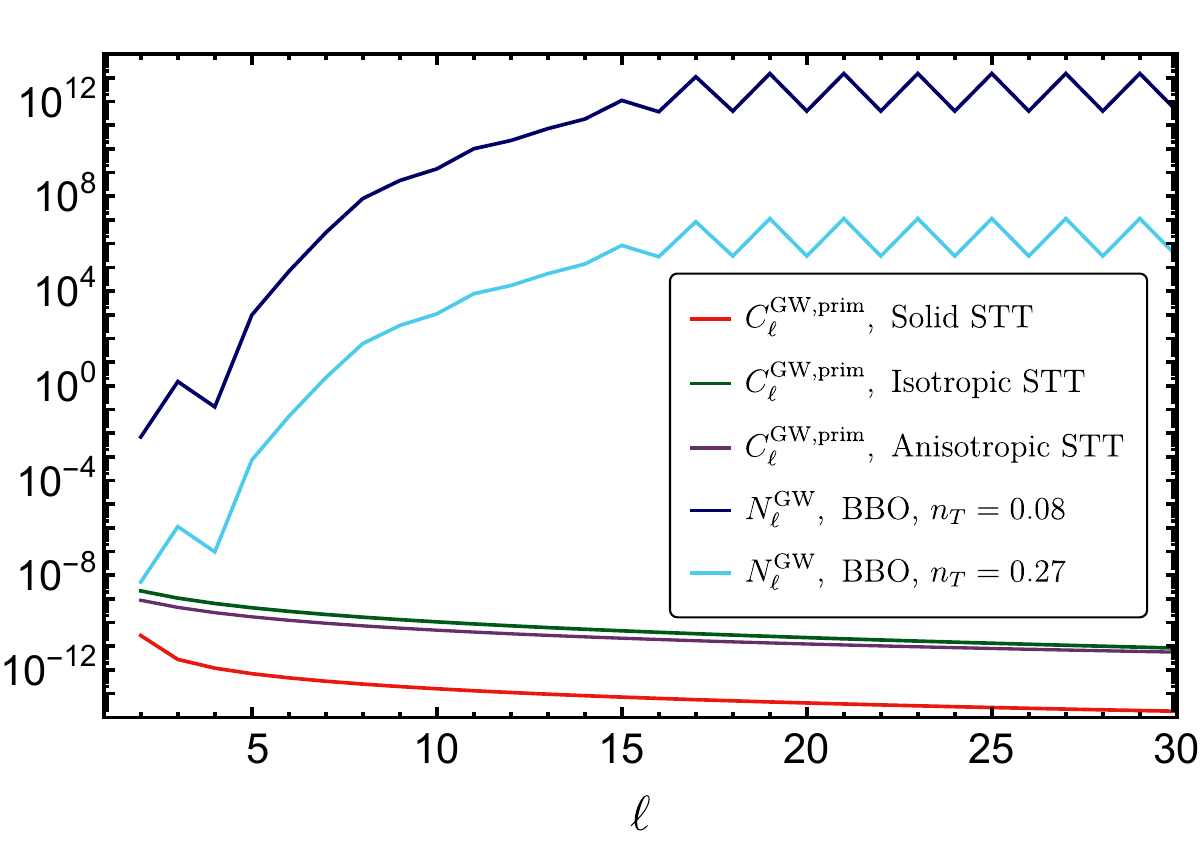}
	\caption{$C_{\ell}^{\text{\tiny{GW,prim}}}$ for the GW anisotropies arising from the solid, isotropic and anisotropic  STT correlations  and the $N_{\ell}^{\text{\tiny{GW}}}$ for BBO with $n_T=0.08,0.27$ and taking $\widetilde{F}_{\rm NL} =1$, $A_S=2.1\times 10^{-9}$. The anisotropic model is introduced in  Section~\ref{stat}. We plot here, for the sake of comparison, the diagonal ($\ell_1 = \ell_2)$ element of the autocorrelation taking $|\lambda_{2M}|=1$.}
	\label{fig:clvnl2}
\end{figure}

In Figure~\ref{fig:clvnl2} we plot the autocorrelation for the solid and isotropic models normalised with $\widetilde{F}_{\rm NL}=1$ and the $N_{\ell}^{\text{\tiny{GW}}}$ for BBO. We see that for solid inflation, with $n_T=0.08$, the signal is always dominated by the noise even with $\widetilde{F}_{\rm NL}\sim 10^3$ (this is very close to its maximum possible value, see \eqref{eq:lambdadef} and Appendix \ref{appA2}). In the case of the isotropic model, a strong signal at the same $n_T$ requires very large non-Gaussianities: $\widetilde{F}_{\rm NL}\sim 10^5$ or larger. On the other hand, for $n_T=0.27$ (a value that is not allowed in the solid inflation model, see Appendix \ref{appA1} ) with $\widetilde{F}_{\rm NL}\gg 1$ the anisotropies are indeed larger than the noise, at least for the first few multipoles. As shown by the violet curve in Fig.~\ref{fig:clvnl2}, a similar behaviour is to be expected for the anisotropic case on which we shall elaborate  in the next section. We conclude that, for the specific set-ups at hand, autocorrelations allow us to probe a rather small region of parameter space where both $n_T$ and $\widetilde{F}_{\rm NL}$ are quite large.

\section{Statistically anisotropic case}
\label{stat}

In spite of the STT non-trivial angular dependence, the solid inflation case is an example of a statistically isotropic set-up. It is then interesting to devote the last part of this manuscript to the study of a statistically anisotropic case. \\
\indent Statistical anisotropy is expected in the kind of models put forward e.g. in \cite{Soda:2012zm,Bartolo:f2F,Emami:2014tpa,Rostami:2017wiy,Bartolo:2015dga}, where the presence of an additional vector field during inflation sets a preferred direction, $\hv$, typically aligned with the vev of the vector field. Solid inflation itself allows for an anisotropic background solution \cite{Bartolo:2013msa,Bartolo:2014xfa}, with features similar to those of the $f(\phi)F^2$ model of \cite{Bartolo:f2F}. While the above references only focus on the scalar bispectrum, $\langle \deltagw_{} \delta^{*\text{\tiny T}}_{}\rangle$ cross-correlations entail instead STT bispectra. The exact form of the STT bispectrum and consequently the cross-correlation would depend in part on the specific inflationary model under consideration. What we wish to perform here is instead a phenomenological analysis, by modelling the form of the STT bispectrum after some of the existing examples of anisotropic scalar bispectra. We assume a non-linearity parameter given by
	\begin{align}
	\fnl(\vq,\vk) = \sum_{LM}\lambda_{LM}(k,q,\hv)Y_{LM}(\hq).    
	\label{eq:anisotropic_fnl}
	\end{align}
	
\begin{figure}
	\centering
	\includegraphics[width=0.7\linewidth]{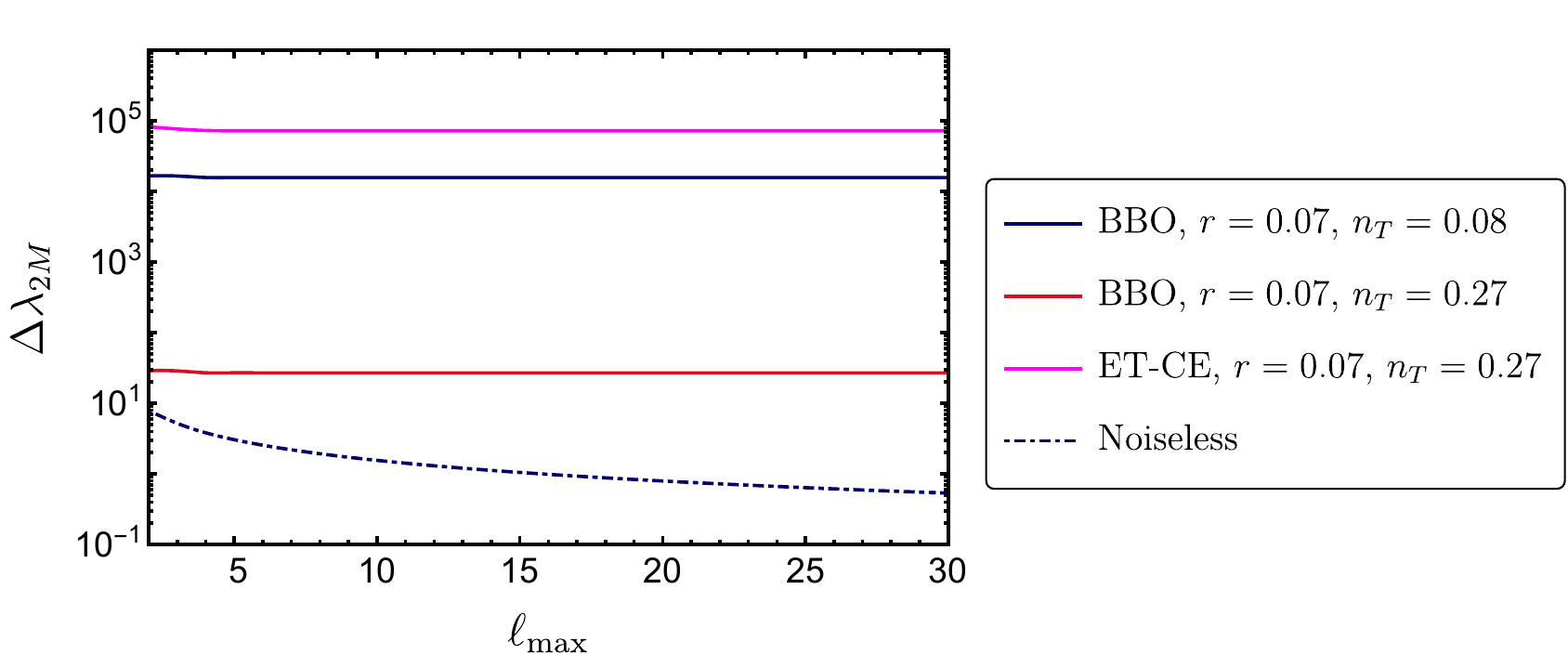}
	\caption{Expected $1\sigma$ error in measuring $\lambda_{2M}$ for BBO, ET-CE, and the idealized noiseless scenario.}
	\label{fig:errorlambda2M}
\end{figure}

Apart from terms of this form, the bispectrum may {also have terms with $(\hv\cdot\hk)$ as well as} contributions such as $(\hv\cdot\hq)(\hv\cdot\hk)(\hk\cdot\hq)$ (as it is the case for e.g. $\langle\zeta^3\rangle$ in \cite{Bartolo:f2F,Ohashi:2013qba,Ohashi:2013mka,Franciolini:2017ktv,Franciolini:2018eno} and $\langle\gamma^3\rangle$ in \cite{Hiramatsu:2020jes}). In addition, there may be anisotropies at the level of the power spectra themselves, e.g. of the form $P_{\gamma}(\vk) = P_{\gamma}(k)(1+\sum_{LM}g_{LM}(k,\hv)Y_{LM}(\hk))$ \cite{Watanabe:2010fh,Bartolo:2017sbu,Obata:2018ilf,Fujita:2018zbr} . In what follows, we focus on the effects of the term in \eqref{eq:anisotropic_fnl}\footnote{{The $(\hv\cdot\hk)$ term can be calculated similarly.} 
As to the contribution from mixed terms such as  $(\hv\cdot\hq)(\hv\cdot\hk)(\hk\cdot\hq)$, which are reminiscent of the structure of the STT bispectrum in solid inflation, we expect those to be suppressed compared to those in Eq.~\eqref{eq:anisotropic_fnl}: as we will show in this section, the solid inflation cross-correlation is smaller compared to the cross-correlation that comes from a term like $(\hv\cdot\hq)^2$.}. As we verify explicitly below, the resulting correlations are statistically anisotropic i.e., there will be non-zero off-diagonal components in $C_{\ell\ell'mm'}^{\text{\tiny GW-T}}$ resulting from the breaking of rotational invariance. For the bispectrum parametrisation \eqref{eq:anisotropic_fnl}, the cross-correlation takes the form
\begin{align}
	\langle{\deltagw_{\ell_1 m_1}\delta^{\text{\tiny T}}_{\ell_2 m_2}}\rangle &= 4\pi i^{\ell_1} \int \frac{d^3q}{(2\pi)^3}Y^*_{\ell_1 m_1}(q)j_{\ell_1}(qd)\sum_{LM}\lambda_{LM}Y_{LM}(\hq)\nonumber\\ 
	& \;\times\frac{4\pi}{5}i^{\ell_2}\int \frac{d^3p}{(2\pi)^3}Y^*_{\ell_2 m_2}(\hp)j_{\ell_2}(pr_{\rm lss})\langle\zeta(\vq)\zeta(\vp)\rangle\nonumber\\
	&=	\frac{2}{5\pi}i^{\ell_1 - \ell_2}\sum_{LM}\left[\int {d^2\hq}\,Y^*_{\ell_1 m_1}(\hq)Y_{LM}(\hq)Y^*_{\ell_2 m_2}(\hq)\int q^2dq\,j_{\ell_1}(qd)j_{\ell_2}(qr_{\rm lss})P_{\zeta}(q)\lambda_{LM} \right]\,.
\end{align}
This simplifies to
\begin{align}
	\langle{\deltagw_{\ell_1 m_1}\delta^{\text{\tiny T}}_{\ell_2 m_2}}\rangle = \frac{2}{5\pi}i^{\ell_1 - \ell_2}\sum_{LM}\left[\int q^2dq\,j_{\ell_1}(qd)j_{\ell_2}(qr_{\rm lss})P_{\zeta}(q)\lambda_{LM} \right](-1)^M h_{\ell_1 \ell_2 L}\tj{\ell_1}{\ell_2}{L}{m_1}{m_2}{-M}\,,
	\label{eq:cross_anisotropic}
\end{align}
where $h_{\ell_{1} \ell_{2} L}$ is defined in Eq.~(\ref{eq:hdef}). This cross-correlation is non-zero for $\ell_1 + \ell_2 + L=\rm even$ and $|\ell_1 - \ell_2| \leq L \leq \ell_1 + \ell_2$, so it will have off-diagonal elements for $L>0$. Such statistically anisotropic correlations are most effectively analysed with the BipoSH formalism \cite{Hajian:2003qq,hajian_cosmic_2004}. The bipolar harmonic coefficients, denoted as $A^{LM,f_1 f_2}_{\ell_1 \ell_2}$, are appropriate combinations of the elements  $\langle{\delta_{\ell_1 m_1}^{f_1}\delta_{\ell_2 m_2}^{f_2}}\rangle$ and are defined as follows. One begins with $\langle{\delta^{f_1}({ \hx}_1)\delta^{f_2}({ \hx}_2)}\rangle$, the real space correlation function of the anisotropies in the observables $f_1,f_2$ corresponding to 2 different directions in the sky ${ \hx}_1$ and ${\hx}_2$. Expanded in the bipolar spherical harmonics this takes the form 
\begin{align}
	\langle{\delta^{f_1}({ \hx}_1)\delta^{f_2}({ \hx}_2)}\rangle = \sum_{\ell_1 \ell_2,LM}A^{LM,f_1 f_2}_{\ell_1 \ell_2}\{Y_{\ell_1}({ \hx}_1)\otimes Y_{\ell_2}({ \hx}_2)\}_{LM}\,,
	\label{eq:bips_def}
\end{align}
where the bipolar spherical harmonics $\{Y_{\ell_1}({ \hx}_1)\otimes Y_{\ell_2}({ \hx}_2)\}_{LM}$ are
\begin{align}
	\{Y_{\ell_1}({ \hx}_1)\otimes Y_{\ell_2}({ \hx}_2)\}_{LM} \equiv \sum_{m_1 m_2}\mathcal{C}_{\ell_1 m_1,\ell_2 m_2}^{LM}Y_{\ell_1 m_1}({ \hx}_1) Y_{\ell_2 m_2}({ \hx}_2)\,,
\end{align}
with 
\begin{align}
	\mathcal{C}_{\ell_1 m_1,\ell_2 m_2}^{LM} = (-1)^{\ell_1-\ell_2+M}\sqrt{2L+1}\tj{\ell_1}{\ell_2}{L}{m_1}{m_2}{-M}
\end{align}
as the Clebsch-Gordan coefficients. Inverting \eqref{eq:bips_def} and doing the angular integration over $({ \hx}_1,{ \hx}_2)$ returns the BipoSH coefficients
\begin{align}
	A^{LM,f_1 f_2}_{\ell_1 \ell_2} = \sum_{m_1 m_2}\mathcal{C}_{\ell_1 m_1,\ell_2 m_2}^{LM}\langle{\delta_{\ell_1 m_1}^{f_1}\delta^{f_2}_{\ell_2 m_2}}\rangle \,.
\end{align}
When statistical isotropy holds, the BipoSH coefficients vanish for $L>0$ and we recover the usual diagonal angular correlations. For the statistically anisotropic case, converting \eqref{eq:cross_anisotropic} to the BipoSH coefficients gives
\begin{align}
  A^{LM, \rm GW-T}_{\ell_1 \ell_2} = i^{\ell_2-\ell_1}\frac{h_{\ell_1 \ell_2 L}}{\sqrt{2L+1}}\left[
{\frac{2}{5\pi}}
    \int q^2dq\,j_{\ell_1}(qd)j_{\ell_2}(qr_{\rm lss})P_{\zeta}(q)\lambda_{LM} \right]\,.
\end{align}
One can then build an unbiased estimator for the BipoSH coefficient
\begin{align}
	\hat{A}^{LM,\rm GW-T}_{\ell_1 \ell_2} \equiv  \sum_{m_1 m_2}\mathcal{C}_{\ell_1 m_1,\ell_2 m_2}^{LM}\deltagw_{\ell_1 m_1}\deltaT_{\ell_2 m_2}\,.
\end{align}
With the Fisher matrix given by
\begin{align}
	F_{\theta_i \theta_j} = \frac{\partial \bm{A}^*}{\partial \theta_i^*}\bm{C}^{-1}_{A^*,A}\frac{\partial^t \bm{A}}{\partial \theta_j}
\end{align}	
and the covariance matrix $\bm{C}$
\begin{align}
	\bm{C}_{A,A} \equiv \left\langle \hat{A}^{LM}_{\ell_1 \ell_2} \hat{A}^{L'M'}_{\ell_1' \ell_2'} \right\rangle_c \,,
\end{align}
one can proceed as in Sec.~\ref{sec5} and calculate the resulting statistical error in the measurement of $\lambda_{LM}$. One finds
\begin{align}
F_{L} = \sum_{\ell_1,\ell_2=\ell_{\rm min}}^{\ell_{\rm max}}\frac{h^2_{\ell_1 \ell_2 L}}{2L+1}\frac{(I_{\ell_1 \ell_2})^2}{C^{\text{\tiny GW}}_{\ell_1}C^{\text{\tiny TT}}_{\ell_2}}\,,
\end{align}
where
\begin{align}
I_{\ell_1 \ell_2} \equiv \frac{2}{5\pi}\int_{q\ll k}q^2dq\,j_{\ell_1}(qd)j_{\ell_2}(qr_{\rm lss})P_{\zeta}(q)\,.
\end{align}
To compare with the solid inflation result, we set $L=2$. The  statistical error is plotted in Figure \ref{fig:errorlambda2M} as a function of $\ell_{\rm max}$ and for different interferometers. For the configuration $|\lambda_{\rm 2\,M}|\sim 10^3,\, n_{T}=0.27$, with BBO one finds 
\bea
\frac{\Delta \lambda_{\rm 2\,M}}{\lambda_{\rm 2\,M}}\sim \mathcal{O}(0.01)\; .
\eea

As in the previous section, we also show the cosmic variance limited error for the noiseless case (see Fig. \ref{fig:errorlambda2M}). Here too we can obtain a simple analytic expression for $\Delta\lambda_{2M}$ if we approximate $d\simeq r_{\rm lss}$, to get
\begin{align}
	F_2 =\frac{3\left(\ell_{\rm max}+3\right)\left(20\ell_{\rm max}^{3}+10\ell_{\rm max}^{2}-33\ell_{\rm max}-12\right)}{1280\pi(4\ell_{\rm max}^{2}+8\ell_{\rm max}+3)}
\end{align}
and 
\begin{align}
	\Delta\lambda_{2M} \equiv F_2^{-1/2} \simeq \frac{16\sqrt{3\pi}}{3\ell_{\rm max}}\,.
	\label{eq:error_anisotropic}
\end{align}

\section{Conclusions}
\label{sec7}

\noindent The detection of primordial gravitational waves would represent a watershed moment for the physics of the early universe. This is true in particular for inflation: for example, in the single-field slow-roll paradigm the tensor to scalar ratio is in one-to-one correspondence with the energy scale at which the acceleration mechanism takes place. In this case then a primordial GW detection would automatically reveal the energy scale at which the cosmological ``collider'' operates, with transformational consequences also for particle physics. The characterisation of  amplitude, scale dependence, chirality and higher-point functions of the GW signal will further enable us to identify important features of multi-field inflationary models.

However, given the presence of the astrophysical stochastic GW background, one should seek additional observables in order to test for the primordial origin of a given signal. A sufficiently large and squeezed scalar-tensor-tensor primordial non-Gaussianity induces in the GW spectrum anisotropies that exceed the contribution from those due to propagation effects. Given their postulated early universe origin, such anisotropies are bound to be correlated with CMB temperature anisotropies: the latter too are sourced by primordial scalar fluctuations. 
Remarkably, given the expected noise for upcoming GW detectors, in certain cases \cite{adshead_multimessenger_2020} the cross-correlation itself is more effective in revealing GW anisotropies than the mere measurement of the GW signal. In this work we have explored the use of cross-correlations as a diagnostic tool for primordial physics in manifold ways.

We first investigated the possibility of using the SGWB and CMB anisotropy cross-correlation to measure the primordial scalar-tensor-tensor bispectrum amplitude $\fnl$ in solid inflation. We find cross-correlations do not provide a  handle on non-Gaussianity for $\fnl\lesssim 10^3$, i.e. those compatible with CMB constraints on solid inflation. This is in contradistinction  to the rather weak condition $\fnl\gtrsim 10$  necessary to achieve a relative error of order $\mathcal{O}(0.1)$ or smaller for a multi-field model with an isotropic bispectrum. 

We also derived the relative error on the non-linearity parameter that would ensue from an idealised, noiseless, survey. We find a relative uncertainty of order $\mathcal{O}(10^{-4})$, $\mathcal{O}(10^{-2}) - \mathcal{O}(10^{-3})$,  in the  $|\tilde{F}_{\rm NL}|\sim 10^3$  configuration  for the isotropic case ($n_T=0.27$) and solid inflation scenario ($n_T=0.08$) respectively.

 Irrespective of the configuration considered, in solid inflation the cross-correlation is suppressed compared to the isotropic case as a result of the non-trivial angular dependence of the STT bispectrum. One should note that the primordial signal (including the STT correlator) supported by these models is far larger than those that typically characterise non-Gaussianities from single field slow-roll inflation. Indeed, the squeezed signal in minimal models is suppressed by powers of $k_L/k_S$, with $k_S$ at interferometer scales and $k_L$ horizon-size. We singled out the $\ell$-dependence (as shown  in Fig.~\ref{fig:comparison}) as a useful handle to distinguish a cross-correlation produced by an STT bispectrum as the one in solid inflation from those originating from primordial STT bispectra as those studied in \cite{adshead_multimessenger_2020} or the ones due to induced anisotropies.

\indent We complemented the focus on a specific model in Section \ref{sec5}  with the phenomenological study of  statistically anisotropic bispectra and cross-correlations in Section \ref{stat}. Our results hold for a wide range of angular behaviours in the non-linearity parameter $F_{\rm NL}$. In order to provide a useful comparison with the solid inflation scenario, we specialised our findings to the $L=2$ case, to show, for the non-linearity parameter $\lambda_{2M}$,  a relative error of $\Delta\lambda_{2M}/\lambda_{2M}\simeq 1\%$ (for BBO with $n_T=0.27, \lambda_{2M}\sim 10^3$). Quite naturally, an even smaller uncertainty characterises the idealised noiseless scenario.

It would be very interesting to consider specific realisations of anisotropic inflation and deliver the sensitivities on the non-linearity parameter associated to upcoming gravitational waves experiments. We leave this to future work.\\
\indent  The recent detection of gravitational wave events has chartered a new course for GW astronomy. Current and next generation GW probes have the potential to usher in a new era also for early universe cosmology. In this work we support, by proving specific examples, the notion that cross-correlations provide a powerful handle on primordial gravitational waves.

\section*{Acknowledgements}
M.\,F. would like to acknowledge support from the
``Atracci\'{o}n de Talento'' grant 2019-T1/TIC-15784. M.\,S. is supported by JSPS KAKENHI Grant Numbers JP19K14718 and JP20H05859. M.\,S. also acknowledges the Center for Computational Astrophysics, National Astronomical Observatory of Japan, for providing the computing resources of Cray XC50.

\appendix

\section{Observational Constraints on Solid Inflation}
\label{appA}

In this section we show that solid inflation can support the values of $(r,n_T)$ considered in Figure \ref{fig:SolidGW}, without violating any of the present constraints  from the CMB. These include the bounds on the scalar spectral index $n_S$ and the tensor-to-scalar ratio $r$ (Appendix \hyperlink{appendix.1.1}{A.1}), and the bounds on primordial non-Gaussianity (Appendix \hyperlink{appendix.1.2}{A.2}).

\subsection{\texorpdfstring{\emph{r} and \emph{n\textsubscript{S}}}{Spectral index and scalar-tensor ratio}}
\label{appA1}

The scalar and tensor power spectra from solid inflation are \cite{endlich_solid_2013},
\begin{align}
	\mathcal{P}_{\zeta}(k) = A_S\left(\frac{k}{k_{ p}}\right)^{n_S-1},\quad 	\mathcal{P}_{\gamma}(k) = A_T\left(\frac{k}{k_{ p}}\right)^{n_T},
\end{align}
where
\begin{align}
	A_S\simeq \frac{H^2_{ p}}{\pi^2 M^2_{\rm Pl}}\frac{1}{8\epsilon c_L^5}\left(\frac{\tau_{ p}}{\tau_{ e}}\right)^{8c_{T}^2\epsilon/3},\quad A_T\simeq \frac{H^2_{ p}}{\pi^2 M^2_{\rm Pl}}\left(\frac{\tau_{ p}}{\tau_{ e}}\right)^{8c_{T}^2\epsilon/3}.
\end{align}
with the scalar and tensor spectral indices and the tensor-to-scalar ratio from solid inflation being,
\begin{align}
n_S - 1 \simeq 2\epsilon c_L^2 - 5s-\eta,\quad n_T \simeq 2\epsilon c_L^2,\quad r \simeq 8\epsilon c_L^5 \,,
\end{align}
where $\epsilon = -\dot{H}/H^2$, $\eta = \dot{\epsilon}/(\epsilon H)$ and $c_L^2$ is the longitudinal sound speed squared with $c_L^2<1/3+2\epsilon/3$ and $s=\dot{c}_L/(c_L H)$. 
The tensor power spectrum is necessarily blue tilted. We show here that the parameter space region considered for solid inflation in Figure \ref{fig:SolidGW} is viable.   \\

CMB observations provide bounds on the scalar and tensor perturbations at large scales, with the most stringent ones coming from  Planck 2018 \cite{Akrami:2018odb}, 
\begin{align}
n_S = 0.9649 \pm 0.0042,\quad r_{0.002}<0.056\,.
\end{align}
Note, however, that the above constraint for $r$ holds when the tensor consistency relation, $n_T = -r/8$, applies. That is not the case for solid inflation. Once the consistency relation is relaxed, the constraints become slightly weaker, $r_{0.01}<0.066$ and $-0.76<n_T<0.52$  \cite{Akrami:2018odb}. This can be converted to a constraint on $r_{0.05}$ using $r_{k'}=r_k(k'/k)^{n_T-n_S+1}$. Further, we set the scalar power spectrum in accordance with Planck, i.e. $A_S = 2.1\times 10^{-9}$. With these constraints in place, we  maximise the magnitude of the tensor power spectrum at interferometer scales identifying the following values for the tensor-to-scalar ratio and spectral index:
 \begin{align}
	r\simeq 0.07\, ,\quad n_T \simeq 0.08\; .
\end{align}

\subsection{Non-Gaussianity}
\label{appA2}
In addition to the above constraints, there are observational bounds on primordial non-Gaussianity, which we present in this subsection. Let us begin with the  solid inflation three-point function for the Bardeen potential $\Phi$ \cite{endlich_squeezed_2014},
\begin{align}
\langle{\Phi_{\vq\to 0}\,\Phi_{\vk}\,\Phi_{-\vk}}\rangle' = -\left(\frac{5}{3}\right)\frac{20}{9}\frac{F_Y}{F}\frac{1}{\epsilon c_L^2}2\mathcal{P}_2(\hq\cdot\hn)P_{\Phi}(q)P_{\Phi}(k)\,.
\end{align}
Given the parametrization
\begin{align}
B_{\Phi}(k_1, k_2, k_3)& = c_2 [\mathcal{P}_2(\hk_1\cdot\hk_2)P_{\Phi}(k_1)P_{\Phi}(k_2)+2\text{ perm.}]\\
& \simeq 2 c_2 \mathcal{P}_{2}(\hq\cdot\hn)P_{\Phi}(q)P_{\Phi}(k)\,,\label{sec}
\end{align}
Planck constraints read $c_2 = -14 \pm 38$  \cite{Akrami:2019izv}.  
In the second line of Eq.~(\ref{sec}) we have taken the squeezed limit $k_1 = q\to 0$ and $k_2=k_3=k$. One finds
\begin{align}
c_2^{\rm solid} &\simeq -\frac{100}{27}\frac{F_Y}{F}\frac{1}{\epsilon c_L^2} = -\frac{100}{27}\frac{F_Y}{F}\frac{2}{n_T} \,.
\end{align}
We require
\begin{align}
	|{c_2^{\rm solid}}|<52,
\end{align}
and with $n_T = 0.08$ one obtains ${F_Y}/{F} \lesssim 0.6$. Since $F_Y/F$ can be freely varied between 0 and 1 \cite{endlich_solid_2013}, this constraint can always be satisfied.
\vspace{0.1cm}

\noindent Next we consider the bounds on tensor non-Gaussianity. In solid inflation one has for the squeezed bispectrum $\langle \gamma_{\vec{q}\to 0}^\lambda \gamma^{\lambda'}_{\vec{k}}\gamma^{\lambda''} _{-\vec{k}}\rangle'$,
\begin{align}
\fnl^{ttt} &\sim \frac{\mathcal{P}_{\gamma}^2}{\mathcal{P}_{\zeta}^2}\cdot \frac{F_Y}{F}\log\left(\frac{k}{aH}\right)   \sim r^2 \cdot \log\left(\frac{k}{aH}\right) \sim \mathcal{O}(0.1) \,, 
\end{align}
where the parameter $\fnl^{ttt}$ is defined as,
\begin{align}
	\fnl^{ttt} = \frac{\langle \gamma_{\vec{q}\to 0}^{\lambda} \gamma^{\lambda'}_{\vec{k}}\gamma^{\lambda''} _{-\vec{k}}\rangle'}{S^{\rm loc}({q, k, k})},\quad S^{\rm loc}({q, k, k}) \equiv \frac{6}{5}(2\pi^2\mathcal{P}_{\zeta})^2\left(\frac{1}{q^3 k^3}+{\text{2 perms.}}\right)
\end{align}
This is again well within the Planck and WMAP limits of $F_{\rm NL}^{ttt} \simeq 290 \pm 180$ and $F_{\rm NL}^{ttt} \simeq 220 \pm 170$ respectively \cite{Shiraishi:2019yux,Ade:2015cva,Shiraishi:2013wua}. In a futuristic experiment like LiteBIRD, one can hope to reach a sensitivity of $F_{\rm NL}^{ttt} \sim \mathcal{O}(1)$ \cite{Shiraishi:2019yux}. 

\indent As to the mixed bispectrum $\langle{\gamma\zeta\zeta}\rangle$, from solid inflation one has in the squeezed limit,
\begin{align}\
\langle{\gamma^{\lambda}_{\vec{q}\to 0}\,\zeta_{\vec{k}}\,\zeta_{-\vec{k}}}\rangle' = -\frac{10}{9}\frac{F_Y}{F}P_{\gamma}(q)P_{\zeta}(k)\frac{1}{c^2_L\epsilon}(\epsilon^{\lambda}_{ij}\hk_i\hk_j)\,.
\end{align}
For the bispectrum parametrization of \cite{Shiraishi:2017yrq},
\begin{align}
\langle{\gamma^{\lambda}_{\vec{q}\to 0}\,\zeta_{\vec{k}}\,\zeta_{-\vec{k}}}\rangle'  = \lim_{q\to 0 } \frac{16\pi^4 g^{tss}A_S^2}{q^3 k^3}\frac{I_{q k k}}{k}\,,
\end{align}
where
\begin{align}
I_{k_1 k_2 k_3} = -(k_1+k_2+k_3)+\frac{k_1 k_2 + k_2 k_3 + k_3 k_1 }{(k_1+k_2+k_3)}+ \frac{k_1 k_2 k_3}{(k_1+k_2+k_3)^2}\,.
\end{align}
Setting $P_{\zeta}(k) = (2\pi^2/k^3)A_S$ and $P_{\gamma}(q) = (2\pi^2/k^3)r \cdot A_S$, one finds
\begin{align}
\lim_{q\to 0 }4g^{tss}\frac{I_{qkk}}{k}\frac{1}{r} = -\frac{10}{9}\frac{F_Y}{F}\frac{1}{c_L^2\epsilon}\,,
\end{align}
and substituting $r = 8\epsilon c_L^5$ one gets
\begin{align}
g^{tss}_{\rm solid} = \frac{40}{27}\frac{F_Y}{F}c_L^3\,.
\end{align}
Using the solid inflation constraints for $F_Y/F$ and $c_L^2$  \cite{endlich_solid_2013}, this parameter is bounded as,
\begin{align}
g^{tss}_{\rm solid}<\frac{40}{27}c_L^3 < 0.095\,,\label{qui}
\end{align}
where the upper bound on $c_L^2$ has been taken to be roughly $c_L^2 \lesssim 1/3$.
Thus the solid inflation prediction for $g^{tss}$ turns out to be compatible with a non-detection as seen in \cite{Shiraishi:2017yrq}. Furthermore, its minimum detectable value in a LiteBIRD level B-mode survey is $g^{tss} \sim \mathcal{O}(1)$ \cite{Domenech:2017kno,Shiraishi:2019yux}, so even future experiments will be not able to set upper bounds at a level which could rule out solid inflation through this kind of measurement.

\section{Computation of the correlations}
\label{appB}

\subsection{Cross-correlation}
\label{appB1}
In this section we present the intermediate steps in the computation of \eqref{eq:cross_result} from the starting point of \eqref{eq:delta_GW}
\begin{align}
\deltagw(k,\hx) = \int_{q\ll k}\frac{d^3q}{(2\pi)^3}e^{i\vq\cdot \vx}\fnl(\vq,\vk)\zeta(\vq) \,.
\end{align}
Upon using the identity for the plane wave decomposition in terms of spherical harmonics,
\begin{align}
e^{i\vq\cdot \vx} = 4\pi\sum_{LM}i^{L} j_{L}(qd)Y_{LM}(\hq)Y^*_{LM}(\hx)
\end{align}
and expanding $\fnl(\vq,\vk)$ using Eqs.~\eqref{equ:fnl}-\eqref{eq:lambdadef} with $\hn=-\hx$ leads to
\begin{align}
\deltagw_{\ell m} = &4\pi\sum_{LM}\sum_{m'}i^{L}\int d^2\hx\,Y^{*}_{\ell m}(\hx)Y^*_{LM}(\hx)Y^*_{2m'}(\hx)\nonumber\\&\times\int_{q\ll k}\frac{d^3q}{(2\pi)^3}j_{L}(qd)Y_{LM}(\hq)Y_{2m'}(\hq)\frac{1}{5}\widetilde{F}_{\rm NL}\zeta(\vq) \,.
\end{align}
The resulting cross correlation is,
\begin{align}
\qop{\deltagw_{\ell_1 m_1}\delta^{*\text{\tiny T}}_{\ell_2 m_2}} = &4\pi\sum_{LM}\sum_{m}i^{L}\int d^2\hx\,Y^{*}_{\ell_1 m_1}(\hx)Y^*_{LM}(\hx)Y^*_{2m}(\hx)\int_{q\ll k}\frac{d^3q}{(2\pi)^3}j_{L}(qd)Y_{LM}(\hq)Y_{2m}(\hq)\nonumber\\&\times\frac{1}{5}\widetilde{F}_{\rm NL}(q,k)
\frac{4\pi}{5}i^{-\ell_2}\int \frac{d^3p}{(2\pi)^3}Y_{\ell_2 m_2}(\hp)j_{\ell_2}(pr_{\rm lss})\langle\zeta(\vq)\zeta^*(\vp)\rangle  
\nonumber\\&= \frac{2}{25\pi}\sum_{LM}\sum_{m}i^{L-\ell_2}\mathcal{G}^{m M m_1}_{2 L \ell_1}\mathcal{G}^{m M m_2}_{2 L \ell_2} \int_{q\ll k} dq\,q^2 \widetilde{F}_{\rm NL}(q,k)j_{L}(qd)j_{\ell_2}(qr_{\rm lss})P_{\zeta}(q) \,,
\end{align}
where $\mathcal{G}^{m_1 m_2 m_3}_{\ell_1 \ell_2 \ell_3}$ is the Gaunt integral,
\begin{align}
\mathcal{G}^{m_1 m_2 m_3}_{\ell_1 \ell_2 \ell_3} &\equiv \int d^2\hq \,Y_{\ell_1 m_1}(\hq)Y_{\ell_2 m_2}(\hq)Y_{\ell_3 m_3}(\hq)\nonumber\\
& = \sqrt{\frac{(2 \ell_1+1)(2\ell_2+1)(2\ell_3+1)}{4\pi}}\tj{\ell_1}{\ell_2}{\ell_3}{0}{0}{0}\tj{\ell_1}{\ell_2}{\ell_3}{m_1}{m_2}{m_3}\,.
\end{align} Using the definition of $G_{\ell_1 \ell_2}$ from \eqref{eq:crossintegral} and $h_{\ell_1 \ell_2 \ell_3}$ from \eqref{eq:hdef} we find,
\begin{align}
\qop{\deltagw_{\ell_1 m_1}\delta^{*\text{\tiny T}}_{\ell_2 m_2}} &= \sum_{LM}\sum_{m}i^{L-\ell_2}\mathcal{G}^{m M m_1}_{2 L \ell_1}\mathcal{G}^{m M m_2}_{2 L \ell_2} G_{L,\ell_2}\nonumber\\
&=\sum_{L}i^{L-\ell_2}h_{2 L \ell_1}h_{2 L \ell_2}G_{L,\ell_2}\sum_{M}\sum_{m}\tj{2}{L}{\ell_1}{m}{M}{m_1}\tj{2}{L}{\ell_2}{m}{M}{m_2}\\
& = \sum_{L}i^{L-\ell_2}h_{2 L \ell_1}h_{2 L \ell_2}G_{L,\ell_2}\frac{\delta_{\ell_1, \ell_2}\delta_{m_1 ,m_2}}{2\ell_1+1}\,,\nonumber
\end{align}
where we have summed over $M,m$ using the following identity,
\begin{align}
\sum_{m,M}\tj{\ell}{L}{\ell_1}{m}{M}{m_1}\tj{\ell}{L}{\ell_2}{m}{M}{m_2} = \frac{\delta_{\ell_1, \ell_2}\delta_{m_1 ,m_2}}{2\ell_1+1}\,.
\label{eq:threejid}
\end{align}
We approximate the comoving distance to the last scattering surface as,
\begin{align}
r_{\rm lss} & =\int_{0}^{z_{\rm lss}}\frac{dz}{H(z)} 
\simeq \frac{1}{H_0}\int_{0}^{1100}\frac{dz}{\sqrt{\Omega_{\rm m}(1+z)^3+\Omega_{\Lambda}}}\\
& \simeq \frac{3.15}{H_0}\,,\nonumber
\end{align}
where the parameters $\Omega_{\rm m}$ and $\Omega_{\Lambda}$ have been set to their Planck values \cite{Aghanim:2018eyx} . We calculate $d$ similarly, by letting $z\to\infty$ since the mode $k$ re-enters the horizon deep in the radiation era and obtain $d\simeq 3.26/H_0$.

\subsection{Autocorrelation}
\label{appB2}
We calculate here the autocorrelation of the solid inflation GW anisotropies. We start from
\begin{align}
\deltagw_{\ell m}=& \frac{4\pi}{5}\sum_{LM,m'}i^L\int d^2\hx\, Y^*_{\ell m}(\hx)Y^*_{LM}(\hx)Y^*_{2m'}(\hx)\nonumber\\ &\times\int \frac{d^3q}{(2\pi)^3}\widetilde{F}_{\rm NL}(q,k)j_{L}(qd)Y_{LM}(\hq)Y_{2m'}(\hq)\zeta(\vec{q})\,.
\end{align}
The autocorrelation is,
\begin{align}
\qop{\deltagw_{\ell_1 m_1} \delta^{*\text{\tiny GW}}_{\ell_2 m_2}}= &\frac{2}{25\pi}\sum_{L_1 M_1 m_1'}\sum_{L_2 M_2 m_2'}i^{L_1-L_2}\mathcal{G}^{m_1 M_1 m_1'}_{\ell_1 L_1 2}\mathcal{G}^{m_2 M_2 m_2'}_{\ell_2 L_2 2}\int q^2 dq\,\widetilde{F}_{\rm NL}^2(q,k)j_{L_1}(qd)j_{L_2}(qd)P_{\zeta}(q)\nonumber\\&\times\int d^2\hq\,Y_{L_1 M_1}(\hq)Y_{2m_1'}(\hq)Y^*_{L_2 M_2}(\hq)Y^*_{2m_2'}(\hq)\,,
\end{align}
where we have written $\qop{\zeta(\vq_1)\zeta^*(\vq_2)}=(2\pi)^3\delta^3(\vec{q}_1-\vec{q}_2)(2\pi^2/q_1^3)A_{\rm S}$.  On further simplification,
\begin{align}
\qop{\deltagw_{\ell_1 m_1} \delta^{*\text{\tiny GW}}_{\ell_2 m_2}}= &\frac{2}{25\pi}\sum_{L_1 M_1,m_1'}\sum_{L_2 M_2,m_2'}i^{L_1-L_2}\mathcal{G}^{m_1 M_1 m_1'}_{\ell_1 L_1 2}\mathcal{G}^{m_2 M_2 m_2'}_{\ell_2 L_2 2}\int q^2 dq\,\widetilde{F}_{\rm NL}^2(q,k)j_{L_1}(qd)j_{L_2}(qd)P_{\zeta}(q)\nonumber\\&\times\sum_{L'M'}\mathcal{G}^{M' M_1 m_1'}_{L' L_1 2}\mathcal{G}^{M' M_2 m_2'}_{L' L_2 2}\,.
\end{align}
By taking the sum over $M_1,m_1'$ and $M_2,m_2'$ and using the identity \eqref{eq:threejid} we obtain,
\begin{align}
\qop{\deltagw_{\ell_1 m_1} \delta^{*\text{\tiny GW}}_{\ell_2 m_2}}&=\sum_{L_1,L_2 }\sum_{L'M'}i^{L_1-L_2}h_{\ell_1 L_1 2}^2 h_{\ell_2 L_2 2}^2\frac{\delta_{\ell_1 L'}\delta_{m_1 M'}}{2\ell_1+1}\frac{\delta_{\ell_2 L'}\delta_{m_2 M'}}{2\ell_2+1}H_{L_1 L_2}\nonumber\\
& = \sum_{L_1,L_2}i^{L_1-L_2}h_{\ell_1 L_1 2}^2 h_{\ell_2 L_2 2}^2\frac{\delta_{\ell_1\ell_2}\delta_{m_1 m_2}}{(2\ell_1+1)^2} H_{L_1 L_2}\,,
\label{eq:clauto_def}
\end{align}
where $H_{L_1 L_2}$ was given in Eq.~(\ref{ref-e}). The geometrical factor $h_{\ell_1 \ell_2 \ell_3}$ was defined in \eqref{eq:hdef} and the sum in \eqref{eq:clauto_def} is now over $L_1,L_2= \ell_1-2,\ell_1,\ell_1+2$. 

\section{Noise Angular Power Spectra of GW Experiments}
\label{appC}
In this section we describe the detector configurations of BBO and ET-CE we work with and briefly review how the corresponding $N_{\ell}^{\text{\tiny{GW}}}$ are calculated. BBO is envisioned to be a system of four LISA like constellations operating in the deci-Hz range. The star configuration of BBO will consist of two constellations rotated $60^{\circ}$ w.r.t each other in the constellation plane. As a later stage of the BBO mission, the other two constellations will be positioned $120^{\circ}$ behind and in front of the star constellation. The individual arm lengths are taken to be $L = 5\times 10^{7}\,\rm m$. In the high frequency range, we consider a ground based detector network consisting of the proposed ET and CE with individual arm lengths of $L = 40\,\rm km$ and $L = 10\,\rm km$ respectively. The sensitivity of ET and CE is quite similar and will be a substantial improvement over other ground based detectors. For the ET-CE coordinates as well as their orientations, we keep the same parameters as in Table 1 of \cite{Alonso:2020rar}.

Here we simply state the basic definitions of the detector quantities involved and the final results for the noise angular power spectra. The detailed calculations can be found in \cite{Alonso:2020rar}. We start with the plane wave expansion of the gravitational background,
\begin{align}
	h_{ij}(\vec{x},t) = \sum_{s=+,\cross}\int df \int d^2\hn\,h_s(f,\hn)e^{i2\pi f (t-\hn\cdot\vx)}e^{s}_{ij}(\hn)\,,
\end{align}
where $s$ denotes the GW polarisation, $f$ is the frequency and $\hn$ the direction of propagation. For unpolarised GW, we can write
\begin{align}
	\langle h_s(f,\hn)h_{s'}(f',\hn')\rangle = \frac{1}{2}\delta(f-f')\frac{\delta^2(\hn-\hn')}{4\pi}\delta_{s s'}I(f,\hn)\,,
\end{align} 
where $I$ is the intensity. The response of a detector at position $\vx$ is the convolution of the detector response tensor $a^{ij}$  with the GW,
\begin{align}
d(f,t) = \int d^2\hn\, \sum_{s}h_{s}(f,\hn)a^{ij}e_{ij}^s(\hn) e^{-i2\pi f\hn\cdot\vx}  + n(t,f)\,,
\end{align}
where $n(t,f)$ is the noise. The noise power spectral density (assuming stationary noise) is defined as 
\begin{align}
	\langle n(f)n(f')\rangle = \frac{1}{2}\delta(f-f')N_f\,.
\end{align}
For a Michelson interferometer with $\hat{u},\hat{v}$ being the unit vectors along the detector arms, the response tensor is given by
\begin{align}
	a^{ij}  = \frac{1}{2}\left[\hat{u}^i\hat{u}^j\mathcal{T}(\hn\cdot\hat u,f)-\hat{v}^i\hat{v}^j\mathcal{T}(\hn\cdot\hat v,f)\right]\,,
\end{align}
where the transfer function $\mathcal{T}(\hn\cdot\hat u,f)$ is 
\begin{align}
	\mathcal{T}(\hn\cdot\hat u,f) = &\frac{1}{2}\bigg[\sinc\left(\frac{\pi f L}{c}(1-\hn\cdot\hat u)\right)e^{-i\frac{\pi f L}{c}(3+\hn\cdot\hat u)}\\
	& + \sinc\left(\frac{\pi f L}{c}(1+\hn\cdot\hat u)\right)e^{-i\frac{\pi f L}{c}(1+\hn\cdot\hat u)}\bigg]\,.
\end{align}
If the intensity $I$ is separable into the product of a direction dependent part and a frequency dependent part, $I(f,\hn) = I_0 (\hn)\xi(f)$, with $\xi(f_{\rm ref})=1$ at a reference frequency $f_{\rm ref}$, its noise angular power spectrum is given by, 
\begin{align}
	N_{\ell}^{-1}\equiv &\frac{1}{2}\sum_{ ABCD}\int df \int dt\, \left(\frac{2\xi(f)}{5}\right)^2 (N^{-1}_f)^{AB}(N^{-1}_f)^{CD}\nonumber\\
	& \times \frac{\sum_m \text{Re}(\mathcal{A}_{BC,\ell m}(t,f)\mathcal{A}^{*}_{DA,\ell m}(t,f))}{2\ell+1}\,,
	\label{eq:Nell_full}
\end{align}
where $N_f^{AB}$ is the noise covariance matrix and $\mathcal{A}_{AB,\ell m}$ the spherical harmonic transform of the antenna pattern for the detector pair AB, the latter being defined as 
\begin{align}
	\mathcal{A}_{AB}(\hn,f) = \underbrace{\frac{5}{8\pi}\left[{\rm Tr}(a^T_A e^+){\rm Tr}(a^T_B e^+)^*+{\rm Tr}(a^T_B e^{\cross}){\rm Tr}(a^T_A e^{\cross})^*\right]}_{\text{overlap function for detector pair AB}} e^{-i2\pi f\hn\cdot(\vx_A-\vx_B)}\,.
\end{align}
We work in the rigid detector approximation (time independent antenna patterns) so that the integral over time can be simplified to 
\begin{align}
	N_{\ell}^{-1}\equiv &\frac{T_{\rm obs}}{2}\sum_{ ABCD}\int df \, \left(\frac{2\xi(f)}{5}\right)^2 (N^{-1}_f)^{AB}(N^{-1}_f)^{CD}\nonumber\\
	& \times \frac{\sum_m \text{Re}(\mathcal{A}_{BC,\ell m}(f)\mathcal{A}^{*}_{DA,\ell m}(f))}{2\ell+1}\,.
	\label{eq:Nell_rigid}
\end{align}
For the ET-CE network we only consider the cross correlations among the ET vertices as well as the ET-CE cross-correlations. All autocorrelations are discarded. As in \cite{Alonso:2020rar}, we make the simplifying assumption that the ET noise correlation matrices are $N_{f}^{AB} = 0.2 \sqrt{N^{A}_f N^{B}_f}$ for $A\neq B$ while the noise will be uncorrelated between the ET-CE vertices. 

In the case of BBO, we only make use of the cross-correlations among detector pairs where the noise can safely be assumed to be uncorrelated, i.e. where $N_f^{AB} = \delta_{AB}N_f^{A}$ which implies that only vertices belonging to different constellations are used. With this choice \eqref{eq:Nell_rigid} can be further simplified to
\begin{align}
	N_{\ell}^{-1}\equiv T_{\rm obs}\sum_{ A,B>A}\int df\, \left(\frac{2\xi(f)}{5}\right)^2 \frac{\sum_m |\mathcal{A}_{AB,\ell m}(f)|^2}{N_f^{A}N_f^B (2\ell+1)}\,.
	\label{eq:Nell_uncorr}
\end{align}
Note that the intensity is related to $\Omegagw$ as follows:
\begin{align}
	\Omegagw(f,\hn) = \frac{4\pi^2 f^3}{3H_0^2}I(f,\hn)\,,
\end{align}
thus the power spectra can be related using
\begin{align}
	C_{\ell}^{\Omegagw}(f) = \left(\frac{4\pi^2 f^3}{3H_0^2}\xi(f)\right)^2 C_{\ell}^{I_0}\,.
\end{align}
The quantity we are interested in, the noise angular power spectrum of the anisotropies $\deltagw$, is given by
\begin{align}
	 N_{\ell}^{\text{\tiny{GW}}}(f) = \frac{N_{\ell}^{\Omega}(f)}{\overline{\Omega}^2_{\text{\tiny{GW}}}(f)}\,,
\end{align}
with $N_{\ell}^{\Omega}$ plotted in Fig.~\ref{Nlps}. We have evaluated this quantity at the reference frequencies $f_{\rm ref}^{\rm BBO}=1\,\rm Hz$  and $f_{\rm ref}^{\rm ET-CE}=63\,\rm Hz$ with the total time of observation taken to be $T_{\rm BBO}=5\,\rm years$ and $T_{\rm ET-CE}=3\,\rm years$.

\bibliographystyle{JHEP}
\bibliography{Cross_SolidInflation}
\end{document}